\documentclass[reprint, floatfix,nofootinbib,amsmath,amssymb,unsortedaddress,superscriptaddress]{revtex4-2}
\usepackage[utf8]{inputenc}
\usepackage{graphicx, cancel, booktabs, tabularx, amsmath, amsfonts, amssymb, bm, hyperref, placeins, float,mathrsfs}
\usepackage[normalem]{ulem}
\usepackage[table]{xcolor}
\usepackage[linesnumbered,ruled,vlined]{algorithm2e}

\newcolumntype{A}{>{\centering\arraybackslash}m{0.9 cm}}
\newcolumntype{B}{>{\centering\arraybackslash}m{0.5 cm}}
\newcolumntype{D}{>{\arraybackslash}m{6 cm}}

\newcolumntype{E}{>{\centering\arraybackslash}m{1.05in}}
\newcolumntype{L}{>{\centering\arraybackslash}m{0.25in}}
\newcolumntype{M}{>{\centering\arraybackslash}m{0.5in}}
\newcolumntype{N}{>{\centering\arraybackslash}m{1.35in}}

\usepackage{csvsimple}
\usepackage{longtable}
\usepackage{booktabs}
\begin{document}
\title{Accelerating Multi-Objective Collaborative Optimization of Doped Thermoelectric Materials via Artiﬁcial Intelligence}

\author{Yuxuan Zeng
% \orcidlink{0009-0004-8954-4377}
}
\affiliation{The Institute of Technological Sciences, Wuhan University, 430072, PR China}

\author{Wenhao Xie}
\affiliation{The Institute of Technological Sciences, Wuhan University, 430072, PR China}

\author{Wei Cao
% \orcidlink{0000-0001-5131-0728}
}
\email{wei\_cao@whu.edu.cn}
\affiliation{The Institute of Technological Sciences, Wuhan University, 430072, PR China}
\affiliation{Key Laboratory of Artiﬁcial Micro- and Nano-Structures of Ministry of Education, School of Physics and Technology, Wuhan University, Wuhan 430072, PR China}

\author{Tan Peng}
\affiliation{Key Laboratory of Artiﬁcial Micro- and Nano-Structures of Ministry of Education, School of Physics and Technology, Wuhan University, Wuhan 430072, PR China}

\author{Yue Hou}
\affiliation{The Institute of Technological Sciences, Wuhan University, 430072, PR China}

\author{Ziyu Wang
% \orcidlink{0000-0001-9718-1263}
}
\email{zywang@whu.edu.cn}
\affiliation{The Institute of Technological Sciences, Wuhan University, 430072, PR China}
\affiliation{Key Laboratory of Artiﬁcial Micro- and Nano-Structures of Ministry of Education, School of Physics and Technology, Wuhan University, Wuhan 430072, PR China}
\affiliation{School of Physics and Microelectronics, Key Laboratory of Materials Physics of Ministry of Education, Zhengzhou University, Zhengzhou 450001, PR China}

\author{Jing Shi}
\affiliation{Key Laboratory of Artiﬁcial Micro- and Nano-Structures of Ministry of Education, School of Physics and Technology, Wuhan University, Wuhan 430072, PR China}

\begin{abstract}
    The thermoelectric performance of materials exhibits complex nonlinear dependencies on both elemental types and their proportions, rendering traditional trial-and-error approaches inefficient and time-consuming for material discovery. In this work, we present a deep learning model capable of accurately predicting thermoelectric properties of doped materials directly from their chemical formulas, achieving state-of-the-art performance. To enhance interpretability, we further incorporate sensitivity analysis techniques to elucidate how physical descriptors affect the thermoelectric figure of merit ($zT$). Moreover, we establish a coupled framework that integrates a surrogate model with a multi-objective genetic algorithm to efficiently explore the vast compositional space for high-performance candidates. Experimental validation confirms the discovery of a novel thermoelectric material with superior $zT$ values in the medium-temperature regime.

\end{abstract}
\date{\today}
\maketitle

\section{Introduction}
\label{sec:Intro}

Thermoelectric materials exploit the Seebeck effect to convert temperature gradients into electrical energy. Compared to conventional energy sources such as nuclear and fossil fuels, thermoelectric technologies offer several distinct advantages, including the absence of moving parts, silent operation, and environmental friendliness~\cite{zhang2014thermoelectric}. In recent years, thermoelectric materials have been widely applied in diverse fields, including health monitoring~\cite{hasan2022health}, waste heat recovery~\cite{bu2022record}, and powering wearable electronics~\cite{siddique2017review}. The performance of a thermoelectric material is typically characterized by the dimensionless figure of merit, $zT$, defined as:  
\begin{align}
    zT = \frac{S^2 \sigma T}{\kappa}
    \label{ztformula}
\end{align}
where $S$ is the Seebeck coefficient, $\sigma$ is the electrical conductivity, $T$ represents the absolute temperature, and $\kappa$ denotes the total thermal conductivity, comprising both lattice ($\kappa_{\rm L}$) and electronic ($\kappa_{\rm e}$) contributions. A higher $zT$ value signifies superior thermoelectric efficiency. 

Widely adopted thermoelectric materials include Bi$_2$Te$_3$-based alloys~\cite{pei2020bi2te3,bano2020room}, which are primarily used at room temperature (below 200 \textcelsius); PbTe and its derivatives~\cite{xiao2018charge,goyal2017first}, optimized for mid-temperature applications (500–600 \textcelsius); and SiGe-based alloys~\cite{toko2024layer,sige}, which demonstrate superior performance in high-temperature environments (above 800 \textcelsius).
The thermoelectric performance of these materials can be significantly enhanced through doping, which enables property optimization beyond the limits of the intrinsic host matrix. Doping strategies in thermoelectrics typically aim to control carrier concentration, engineer the electronic band structure, and reduce lattice thermal conductivity. Common approaches include donor and acceptor doping to modulate electron and hole populations~\cite{yang2018enhancing}; band convergence to increase the density-of-states effective mass and improve charge transport~\cite{lv2014enhanced,pei2012band}; and defect engineering to strengthen low-frequency phonon scattering and suppress thermal conductivity~\cite{zheng2021defect}.
Although the synthesis of doped thermoelectric materials is well-established, the selection of host matrices, dopant species, and their concentrations remains largely empirical. As a result, the discovery of high-performance compositions still relies heavily on time-consuming and costly trial-and-error processes.
Furthermore, the computational prediction of thermoelectric properties in doped systems poses considerable challenges. For density functional theory (DFT) calculations, the need to construct large supercells~\cite{lany2009accurate,kang2022advances} and account for band structure reconstruction~\cite{chen2012effect} leads to substantial computational overhead. Similarly, molecular dynamics (MD) simulations face difficulties such as the lack of accurate force fields~\cite{fang2014recent}, limited accessible timescales~\cite{millett2006molecular}, and other resource constraints.
\begin{figure*}
    \centering
    \includegraphics[scale=0.1]{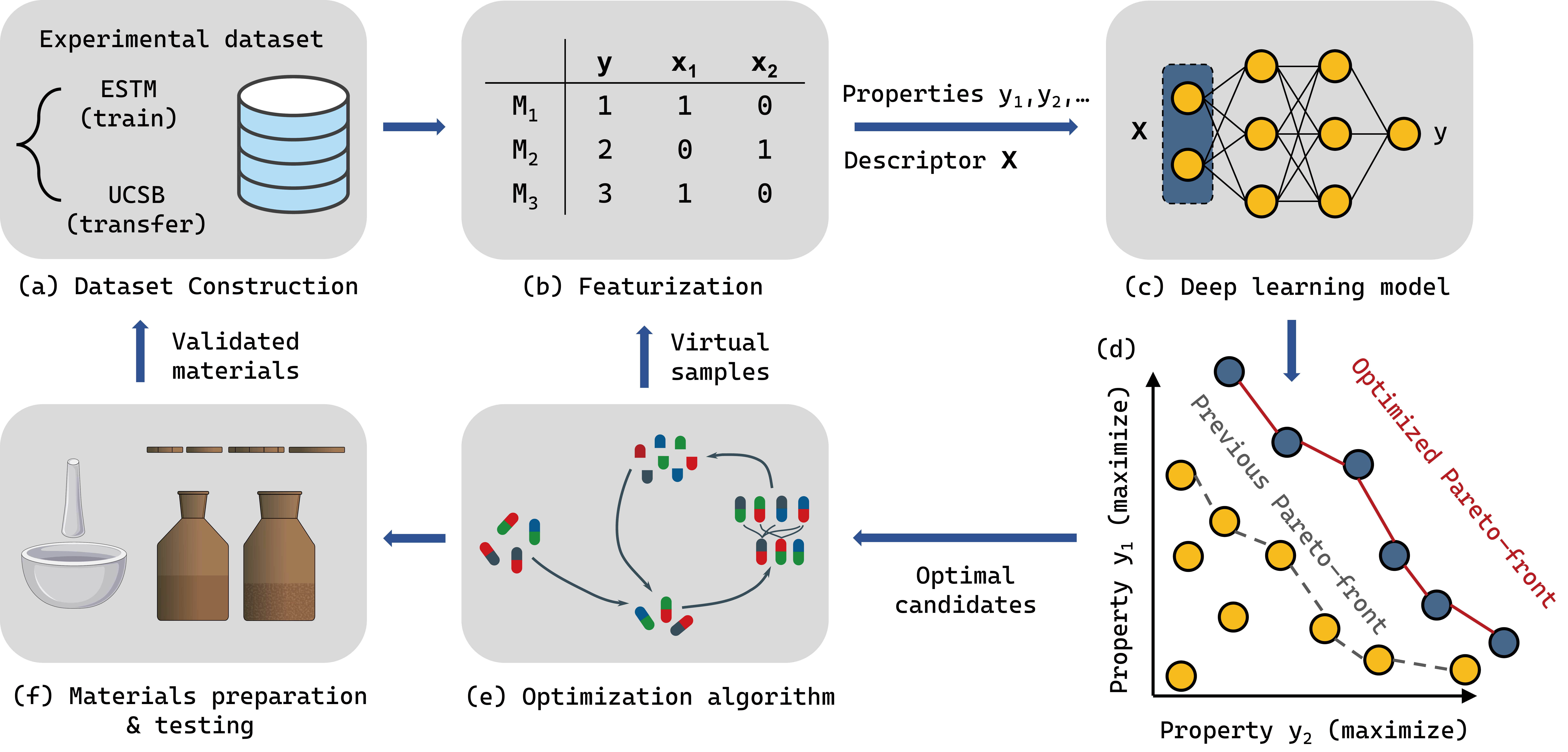}
    \caption{The proposed deep learning–optimization algorithm coupled framework is designed for multi-objective collaborative optimization in the design of thermoelectric materials with high $zT$ values. (a) Two datasets are employed separately for deep learning training and transfer learning testing, both derived from experimental sources. (b) Digitized descriptors are used to represent different materials. (c) Deep learning is utilized to construct the mapping between descriptors and thermoelectric performance. (d) The Pareto-front is adopted to select a subset of candidate materials with optimal performance from the dataset. (e) The optimization algorithm is applied to explore promising virtual samples from the vast compositional space. (f) Upon completion of the optimization iterations, the most promising candidate(s) are selected for experimental characterization and mechanistic investigation.
    }
\end{figure*}

In recent years, machine learning and deep learning have emerged as powerful data-driven methodologies for addressing complex challenges in thermoelectric materials research, such as predicting power factors~\cite{ZENG2025102627,lantunes} and modeling thermal conductivity~\cite{luo2023predicting}.
To date, the majority of AI-assisted materials studies have concentrated on intrinsic compounds. This is largely attributed to the fact that intrinsic materials possess well-defined crystal and electronic band structures, rendering them more tractable for characterization and computational modeling. Additionally, the availability of large-scale, open-access databases—such as the Materials Project~\cite{mpdatabase} and AFLOW~\cite{curtarolo2012aflow}—greatly facilitates data acquisition by providing abundant, high-quality training data. Furthermore, for intrinsic systems, computational methods like density functional theory (DFT) and molecular dynamics (MD) are not only practical but also reliable for validating machine learning predictions, and in some cases, for enhancing model performance through active learning strategies~\cite{sheng2020active}.

In the prediction of doped materials’ properties, three primary challenges have been identified~\cite{dopnet}: (1) the absence of crystal structure information; (2) the dilution of doping effects—subtle variations in dopant concentration are often difficult to capture as meaningful distinguishing features; and (3) strong nonlinear effects—changes in dopant species and concentrations can have significant and highly nonlinear impacts on material properties.
In the context of thermoelectric transport properties in doped systems, several notable efforts have been made. Gyoung et al. introduced DopNet\cite{dopnet}, a deep learning framework specifically developed for predicting the thermoelectric performance of doped materials. In parallel, Parse et al.\cite{web} developed a user-interactive platform~\cite{weblink}, enhancing accessibility and usability for researchers.
While these studies offer valuable insights, several limitations persist:
\begin{itemize}  
\item Suboptimal model accuracy, manifested in overfitting to the training set~\cite{web} and limited extrapolation capability~\cite{dopnet}.  
\item Dependence on experimental data for feature construction~\cite{web}.  
\item Evaluation of extrapolation ability solely through model transfer across different datasets, lacking experimental validation~\cite{dopnet,na2022public}.  
\item Absence of inverse material design considerations~\cite{dopnet,na2022public,web}.
\end{itemize}

To enable efficient exploration of doped thermoelectric materials, it is essential not only to address the aforementioned challenges, but also to account for the vast combinatorial design space involving host matrices, dopant species, and their concentrations.
In related fields such as high-entropy alloy design, optimization strategies like genetic algorithms~\cite{guo2021machine,WEN2024} and Bayesian optimization~\cite{khatamsaz2022multi} have been successfully applied to inverse material design. However, in the context of thermoelectric materials, the electrical transport parameters—namely the Seebeck coefficient ($S$) and electrical conductivity ($\sigma$)—are inherently antagonistic, as improvements in one often lead to deterioration of the other. This trade-off is further compounded by the effects of electron–phonon coupling~\cite{heid2017electron}, which intricately link electrical and thermal transport processes.
\begin{figure*}
    \centering
    \includegraphics[scale=0.65]{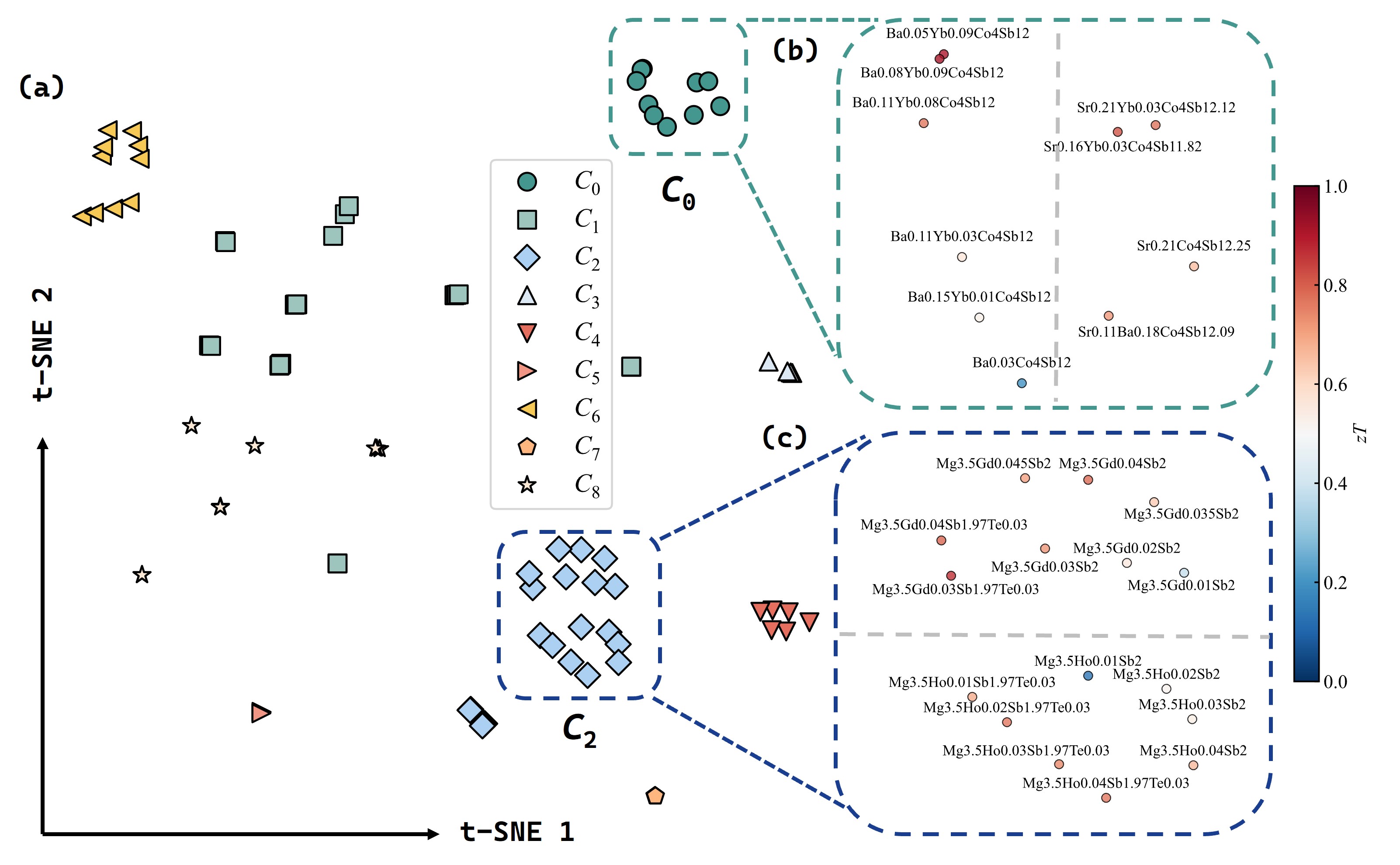}
    \caption{$k$-means clustering and t-SNE 2D visualization. (a) All samples at 600~K are divided into 9 clusters, from $C_0$ to $C_8$, distinguished by different scatter plot markers. (b) and (c) show local zoom-ins of $C_0$ and $C_2$, respectively, with color mapping indicating the $zT$ values of the samples.}
    \label{tsne-non}
\end{figure*}

This work aims to address the inefficiency of trial-and-error strategies in thermoelectric material discovery. We propose a wavelet-based feature enhancement method that extracts both inter- and intra-system variations from chemical formulas. Based on this, we develop the \underline{Wave}let-enhanced \underline{T}hermo\underline{e}lectric \underline{Net}work (WaveTENet), a deep learning model that simultaneously predicts five thermoelectric performance criteria: $S$, $\sigma$, PF, $\kappa$, and $zT$. Validated on multiple datasets, WaveTENet achieves state-of-the-art performance. With the integration of the \underline{SH}apley \underline{A}dditive ex\underline{P}lanation (SHAP) framework, we conduct sensitivity analysis to interpret the role of physical descriptors in determining $zT$. Coupled with the \underline{N}on-dominated \underline{S}orting \underline{G}enetic \underline{A}lgorithm \underline{III} (NSGA-III), the framework enables inverse design by optimizing compositions based on model feedback. We identify several high-$zT$ candidates, one of which is experimentally validated. Although this study focuses on thermoelectrics, the methodology is applicable to broader material domains including high-entropy alloys, superconductors, and magnetic materials.

\section{Results}

\subsection{Nonlinear Effects of Doping on thermoelectric Performance}
\label{nonlinear}

The influence of doping on thermoelectric materials is intrinsically nonlinear, as even slight variations in elemental type or composition can lead to abrupt and pronounced changes in thermoelectric performance~\cite{al2024effect,bernazzani2023bipolar}. To validate this perspective, we selected multiple samples at 600~K from the dataset for $k$-means clustering. The features used for material characterization are detailed in Sec.~\ref{feat}; these features are not only employed for clustering but also serve as input for subsequent deep learning model training. We determined the number of clusters as $k = 9$ using the Elbow method~\cite{humaira2020determining}.

In Fig.~\ref{tsne-non}(a), all samples are grouped into nine clusters, among which clusters $C_0$ and $C_2$ tend to exhibit higher $zT$ values. As shown in Fig.~\ref{tsne-non}(b), cluster $C_0$ is primarily composed of derivatives of Co$_4$Sb$_{12}$, a cobalt antimonide with a skutterudite structure~\cite{skutterudite}. The left subgroup is dominated by Ba- and Yb-doped samples, whereas the right subgroup mainly consists of Sr-doped compositions with additional Sb incorporation.
Although most samples in $C_0$ demonstrate relatively high $zT$ values, Ba$_{0.03}$Co$_4$Sb$_{12}$ notably exhibits a significantly lower $zT$ of 0.211. Similarly, Fig.~\ref{tsne-non}(c) provides a magnified view of cluster $C_2$, which mainly comprises derivatives of Mg$_{3}$Sb$_2$~\cite{shi2018advances}. Notably, $C_2$ is split into two subgroups: the upper subgroup is predominantly doped with Gd and Te, while the lower subgroup primarily features Ho and Te doping. Despite both Mg$_{3.5}$Ho$_{0.01}$Sb$_2$ ($zT = 0.159$) and Mg$_{3.5}$Gd$_{0.01}$Sb$_2$ ($zT = 0.431$) belonging to this cluster, their $zT$ values remain lower than those of other high-performing samples.

Such nonlinear behavior poses a significant challenge for supervised learning. When predicting the performance of an unknown sample, reference values derived from similar compositions within the same system in the training data are inherently limited. Both over-reliance on and complete disregard for structurally similar materials may negatively affect model predictions. Therefore, it is essential that both feature engineering and deep learning model design explicitly account for this phenomenon. On the one hand, the influence of similar samples should be carefully mitigated; on the other hand, sufficient nonlinear fitting capacity must be ensured through deep network architectures.

\subsection{Multi-Source Feature Fusion and Wavelet-Based Enhancement}
\label{feat}
In materials informatics, mainstream feature construction methods include empirical features, structural features~\cite{jarvis,himanen2020dscribe,pham2017machine}, embedding vectors~\cite{zhou2018learning}, and elemental composition statistics.

Empirical features refer to variables obtained through experiments or DFT calculations that have a theoretical correlation with the target property. Machine learning models based on such features often achieve high accuracy with ease; however, acquiring these features can be challenging, making them unsuitable for high-throughput screening. Structural features describe the electronic configuration within a crystal and are represented as vectors, but characterizing doped systems in this manner is not straightforward. Mat2Vec~\cite{tshitoyan2019unsupervised} is a representative embedding vector approach that draws inspiration from word embeddings in natural language processing. However, such features generally lack interpretability, as the individual dimensions of the vector do not have explicit physical meanings.  

To efficiently and accurately represent doped systems, we preliminarily select Magpie~\cite{ward2016general} as one of the feature sources. Magpie analyzes a given chemical formula and extracts interpretable physical, chemical, electronic, and ionic properties—such as melting point, covalent radius, and electronegativity—for each element. It then computes statistical measures such as the mean and variance based on elemental composition. This process does not rely on experimental data or structural parameters, enabling the rapid construction of sample features.

System-Identified Material Descriptor (SIMD)~\cite{na2022public} has demonstrated remarkable performance in predicting the properties of doped thermoelectric materials. The SIMD vector consists of four components:
\begin{align}
    \mathcal{S}=\mathbf{x}_{s} \oplus \mathbf{c}_{s} \oplus \mathbf{w}_{s}^{(K)} \oplus \mathbf{w}_{s}^{(K)}
    \label{simdeq}
\end{align}
where $\mathbf{x}_{s}$ is a 100-dimensional vector encoding the fractional composition of elements from H to Fm in the chemical formula. $\mathbf{c}_{s}$ is a conditional vector incorporating synthesis conditions, which, in this work, only considers temperature conditions. $\mathbf{w}_{s}^{(K)}$ represents the distance-weighted sum of the system vectors of input chemical composition $s$ in the system identification process, $\mathbf{w}_{s}^{(K)}$ denotes the distance-weighted sum of the target statistical vectors for the selected $K$ nearest material clusters, while the symbol ``$\oplus$'' signifies vector concatenation. A more detailed explanation can be found in Sec.~\ref{desc-method}. 
\begin{figure*}
    \centering
    \includegraphics[scale=0.067]{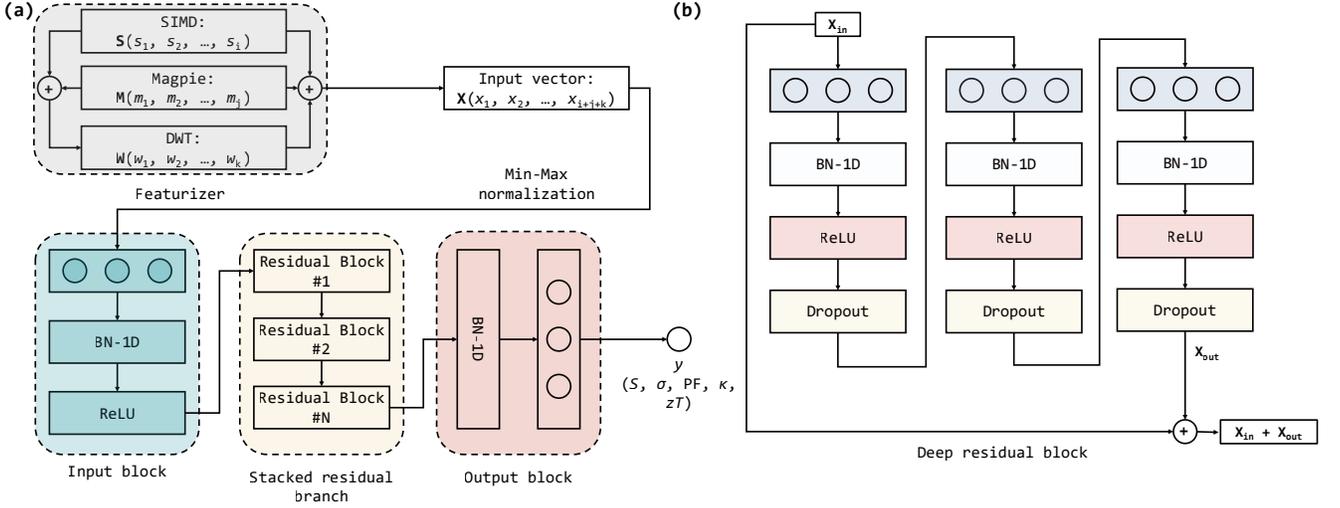}
    \caption{Architecture of \textbf{WaveTENet}. (a) The network is composed of three main components: the input block, the stacked residual branch, and the output block. (b) The structure of the deep residual module. The modules depicted as rectangles with embedded circles represent linear layers.}
    \label{fig:net}
\end{figure*}

SIMD aggregates training samples similar to a predictive sample to form an anchor space. Using $k$-nearest neighbor ($k$NN) method, information from the $k$ most similar known clusters is embedded into $\mathbf{w}_{s}^{(K)} $ and $\mathbf{w}_{s}^{(K)} $. This approach effectively constrains the predicted properties of new samples within a reasonable domain based on prior knowledge. Fundamentally, this is a process of ``seeking similarity.''  
However, considering the nonlinear effects discussed in Sec.~\ref{nonlinear}, particularly in cases like Mg$_{3.5}$Ho$_{0.01}$Sb$_2$, such an anchor space may instead introduce adverse effects.

The variations in elemental composition and fraction can be captured by Magpie descriptors. However, when the fractional changes are minimal, the corresponding variations in Magpie descriptors are also subtle, making them insufficient to mitigate the over-reliance on the anchor space. To address this, we apply Discrete Wavelet Transform (DWT)~\cite{bentley1994wavelet,sundararajan2016discrete} to the features and incorporate the wavelet coefficients as supplementary descriptors, thereby achieving a ``preserving differences'' approach. DWT can be used to provide local slope estimation in time series data to assess similarity~\cite{xiaoboxiangsiqifa}. Inspired by this, we aim to amplify the differences among materials within the same doping system. 

In our case, we concatenated the SIMD vector $\mathcal{S}$ and the Magpie vector $\mathcal{M}$, treating them as a discrete-time signal. Using the Haar wavelet~\cite{lepik2014haar}, the original feature signal was decomposed into approximation coefficients ($\mathcal{A}$) and detail coefficients ($\mathcal{D}$). The former represented the main trend (smooth component) of the signal, emphasizing differences between different doping systems, while the latter captured rapid variations (detail component), distinguishing subtle differences within the same doping system. After concatenating the wavelet coefficients with the initial features, we obtained the final feature representation for model input.
\begin{align}
    \mathcal{X}=\mathcal{S} \oplus \mathcal{M} \oplus \mathcal{A} \oplus \mathcal{D}
    \label{input-vec}
\end{align}

Here, we employ cosine similarity (CS) to evaluate the distinguishability of samples at the feature level.
\begin{table}[h]
    \centering
    \caption{Cosine similarity in $C_2$ cluster. $\mathcal{S} \oplus \mathcal{M}$ represents the computation result using only SIMD and Magpie features, while $\mathcal{S} \oplus \mathcal{M} \oplus \mathcal{A} \oplus \mathcal{D}$ denotes the computation after incorporating wavelet coefficients. The arrows indicate the changes in cosine similarity of the samples before and after introducing DWT transformation.}
    \begin{tabular}{lccc}
        \toprule
        \textbf{Formula} & $zT$ & $\mathcal{S} \oplus \mathcal{M}$ & $\mathcal{S} \oplus \mathcal{M} \oplus \mathcal{A} \oplus \mathcal{D}$ \\
        \midrule
        Mg$_{3.5}$Ho$_{0.01}$Sb$_2$ & 0.1588 & - & - \\
        Mg$_{3.5}$Ho$_{0.02}$Sb$_2$ & 0.5902 & 0.9831 & 0.9759 $\downarrow$ \\
        Mg$_{3.5}$Ho$_{0.03}$Sb$_2$ & 0.6094 & 0.9305 & 0.9032 $\downarrow$ \\
        Mg$_{3.5}$Ho$_{0.04}$Sb$_2$ & 0.7487 & 0.8480 & 0.7961 $\downarrow$ \\
        Mg$_{3.5}$Gd$_{0.01}$Sb$_2$ & 0.4314 & 0.7404 & 0.7860 $\uparrow$ \\
        Mg$_{3.5}$Gd$_{0.02}$Sb$_2$ & 0.6203 & 0.6491 & 0.6899 $\uparrow$ \\
        Mg$_{3.5}$Gd$_{0.03}$Sb$_2$ & 0.8175 & 0.5183 & 0.5483 $\uparrow$ \\
        Mg$_{3.5}$Gd$_{0.035}$Sb$_2$ & 0.7121 & 0.4468 & 0.4713 $\uparrow$ \\
        Mg$_{3.5}$Ho$_{0.01}$Sb$_{1.97}$Te$_{0.03}$ & 0.7781 & 0.4270 & 0.4505 $\uparrow$ \\
        Mg$_{3.5}$Ho$_{0.02}$Sb$_{1.97}$Te$_{0.03}$ & 0.8634 & 0.3894 & 0.4087 $\uparrow$ \\
        Mg$_{3.5}$Ho$_{0.03}$Sb$_{1.97}$Te$_{0.03}$ & 0.8395 & 0.3373 & 0.3465 $\uparrow$ \\
        Mg$_{3.5}$Gd$_{0.045}$Sb$_2$ & 0.7984 & 0.3093 & 0.3251 $\uparrow$ \\
        Mg$_{3.5}$Ho$_{0.04}$Sb$_{1.97}$Te$_{0.03}$ & 0.8632 & 0.2761 & 0.2737 $\downarrow$ \\
        Mg$_{3.5}$Gd$_{0.04}$Sb$_2$ & 0.8884 & 0.3763 & 0.3960 $\uparrow$ \\
        Mg$_{3.5}$Gd$_{0.03}$Sb$_{1.97}$Te$_{0.03}$ & 0.9778 & 0.0056 & 0.0566 $\uparrow$ \\
        Mg$_{3.5}$Gd$_{0.04}$Sb$_{1.97}$Te$_{0.03}$ & 0.8964 & -0.0947 & -0.0519 $\uparrow$ \\
        Mg$_{3.07}$Sb$_{1.5}$Bi$_{0.48}$Se$_{0.02}$ & 0.9604 & -0.7134 & -0.7088 $\uparrow$ \\
        Mg$_{3.07}$Sb$_{1.5}$Bi$_{0.47}$Se$_{0.03}$ & 0.9032 & -0.7205 & -0.7255 $\downarrow$ \\
        Mg$_{3.07}$Sb$_{1.5}$Bi$_{0.46}$Se$_{0.04}$ & 0.9064 & -0.7235 & -0.7347 $\downarrow$ \\
        Mg$_{3.07}$Sb$_{1.5}$Bi$_{0.45}$Se$_{0.05}$ & 0.8477 & -0.7225 & -0.7362 $\downarrow$ \\
        Mg$_{3.07}$Sb$_{1.5}$Bi$_{0.44}$Se$_{0.06}$ & 0.8254 & -0.7176 & -0.7305 $\downarrow$ \\
        \botrule
    \end{tabular}
    \label{tab:cosine_similarity}
\end{table}
Taking the $C_2$ cluster shown in Fig.~\ref{tsne-non}(a) and (c) as an example, Table~\ref{tab:cosine_similarity} revealed that for Mg$_{3.5}$Ho$_{0.01}$Sb$_2$, the CS of samples within the same doping system, Mg$_{3.5}$Ho$_{x}$Sb$_2$, generally decreased significantly. In contrast, samples that originally exhibited greater differences (CS $<0.8$) tended to show an increase in CS after the introduction of DWT. This suggested that during model training, the inclusion of DWT broadened the model's perspective, enabling it to incorporate a more diverse set of samples into decision-making rather than focusing on a limited number of similar materials within the same doping system. Consequently, this enhanced the model's ability to capture the nonlinear effects of the target properties more effectively.
\subsection{Architecture of WaveTENet}
We designed WaveTENet to achieve multi-objective regression of thermoelectric transport properties for doped material compositions. WaveTENet consists of four functional blocks: \texttt{Featurizer}, \texttt{Input block}, \texttt{Deep residual block}, and \texttt{Output block}. 

The \texttt{Featurizer} directly generates the corresponding SIMD vector $\mathcal{S} \in \mathbb{R}^{n \times d_{\text{SIMD}}}$ and Magpie vector $\mathcal{M} \in \mathbb{R}^{n \times d_{\text{Magpie}}}$ based on the input material formula and concatenates them. Treating $\mathcal{S} \oplus \mathcal{M}$ as a discrete signal sequence, a first-level Haar wavelet transform is applied to obtain the low-frequency coefficient vector $\mathcal{A} \in \mathbb{R}^{n \times (d_{\text{SIMD}}+d_{\text{Magpie}})/2}$ and the high-frequency coefficient vector $\mathcal{D} \in \mathbb{R}^{n \times (d_{\text{SIMD}}+d_{\text{Magpie}})/2}$. The variables $n$ and $d$ represent the number of samples in the training data (or a single batch) and the dimensionality of the features, respectively. The number of wavelet coefficients obtained from a single-level Haar wavelet transform is half the length of the original signal~\cite{leavey2003introduction}. In our case, the specific dimension of each vector are:
\begin{align}
    \begin{cases}
    d_{\text{SIMD}} = 219, \\
    d_{\text{Magpie}} = 255, \\
    d_{\text{A}} = d_{\text{D}} = 237.
    \end{cases}
\end{align}
Eq.~(\ref{input-vec}) is the output of \texttt{Featurizer}, denoted as $\mathcal{X}$, which will be read by the \texttt{Input block} for further processing after min-max normalization.  

The task of the \texttt{Input block} is to perform fundamental processing on the input features. This block consists of three interconnected components: a linear layer, a batch normalization (BN) layer, and an activation function layer. 

We set the input and output dimensions of the linear layer to be identical, thereby preserving the dimensionality of the initial feature vector $\mathbf{x}$. This choice is motivated by the fact that, in our study, the feature differences among samples within the same doping system are inherently subtle. Reducing dimensionality through a linear layer would further obscure these distinctions, potentially compromising the model’s ability to capture meaningful variations.  
The BN layer computes the mean and standard deviation for each mini-batch and normalizes the data:
\begin{align}
    \mathcal{\hat{X}} = \frac{\mathcal{X} - \mathcal{\mu}}{\mathcal{\sigma}}
\end{align}
where $\mu$ is the feature mean vector, and $\sigma$ is the feature standard deviation vector. The significance of BN lies in:
\begin{itemize}
    \item Ensuring that data from different batches have similar distributions, thereby improving the model's generalization ability~\cite{furusho2020theoretical}.
    \item Addressing the issue of ``internal covariate shift.''~\cite{ioffe2015batch}
\end{itemize}

Unlike conventional affine transformations, all linear layer outputs in WaveTENet are first processed by BN before being passed through the ReLU activation function. This design primarily serves to alleviate the vanishing gradient issue associated with ReLU~\cite{daneshmand2020batch}.  

The \texttt{deep residual block} consists of multiple residual blocks connected in series, where each residual block is composed of three sequentially stacked basic modules, each following the structure of ``Linear layer $\to$ BN $\to$ ReLU $\to$ Dropout layer.'' Experimental results indicate that using six stacked residual blocks achieves an optimal balance between model performance and training efficiency.
Dropout is implemented by randomly setting a fixed proportion of neuron outputs to zero during each training iteration, thereby preventing overfitting and enhancing generalization ability~\cite{gal2016theoretically}. Moreover, to equip the model with strong nonlinear learning capabilities, WaveTENet incorporates multiple stacked linear layers. To mitigate the risk of gradient explosion or vanishing gradients, we adopt residual connections inspired by the ResNet~\cite{he2016deep}.  

The structure of the \texttt{output block} is significantly simpler, consisting only of a BN layer and a linear layer. Following the \texttt{output block}, a single neuron is fully connected to the linear layer, serving as the output head of the model.  
\subsection{Model Training \& Evaluation}

Experimentally Synthesized Thermoelectric Materials (ESTM)~(\url{https://github.com/KRICT-DATA/SIMD}) is an open-source thermoelectric material dataset constructed via text mining~\cite{na2022public}, encompassing both doped and intrinsic compounds with thermoelectric performance metrics recorded over a temperature range from as low as 10~K to as high as 1,275~K, comprising up to 5,205 entries. In our study, we excluded intrinsic compounds from the ESTM dataset, as their thermoelectric properties require additional consideration of carrier concentration~\cite{lantunes}. In contrast, doping inherently serves as a direct means of modulating carrier concentration~\cite{dopecarrier}, which can be reflected in the chemical formula, thereby eliminating the need for an additional carrier-related parameter in doped materials.

During training, we incorporated $\ell_2$ regularization into WaveTENet to mitigate overfitting. Given that most PF values fall within the range of $10^{-5}$ to $10^{-3}$, numerical precision issues in floating-point operations could potentially affect model fitting~\cite{gupta2015deep}. To address this, we applied a scaling factor of $10^5$ to PF during data loading. In contrast, $\sigma$ exhibits a broad numerical range, spanning from a minimum of 0 to a maximum on the order of $10^7$. Such a large-scale variation in values can disrupt deep learning model training; hence, we introduced a scaling factor of $10^{-3}$ to $\sigma$ to compress the distribution of the target property.
\begin{figure*}
    \centering
    \includegraphics[scale=0.35]{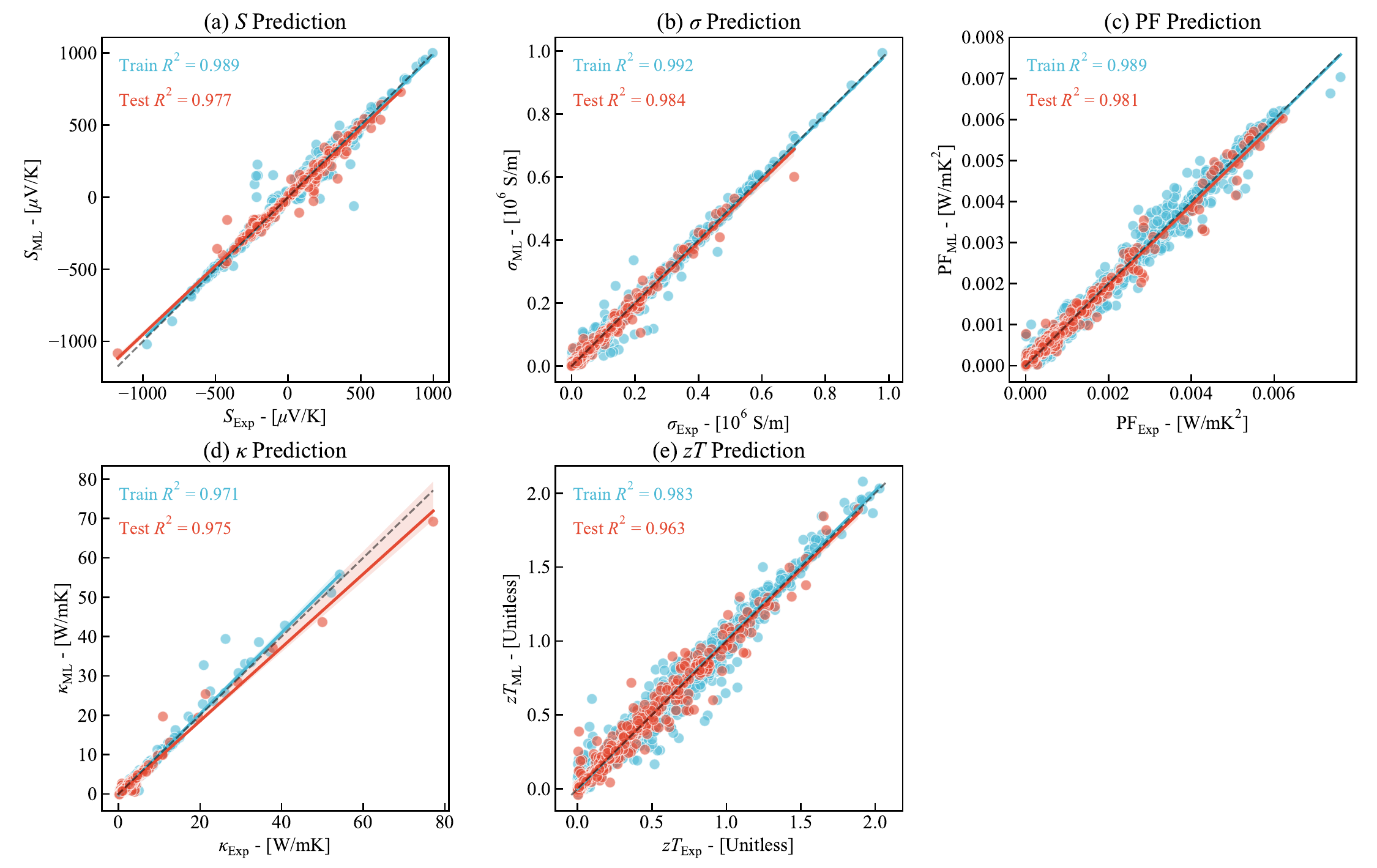}
    \caption{The scatter plots of WaveTENet’s predictions for $S$, $\sigma$, PF, $\kappa$, and $zT$ are presented. Each point in the plots corresponds to a sample from the dataset, with the $x$-axis representing the ground truth (experimentally measured values) and the $y$-axis indicating the model's predicted values. The black dashed diagonal line denotes the ideal case where predictions perfectly match the true values. The solid blue and red lines represent the regression fits for the training and test sets, respectively.}
    \label{fig:performance}
\end{figure*}
\begin{figure*}
    \centering
    \includegraphics[scale=0.35]{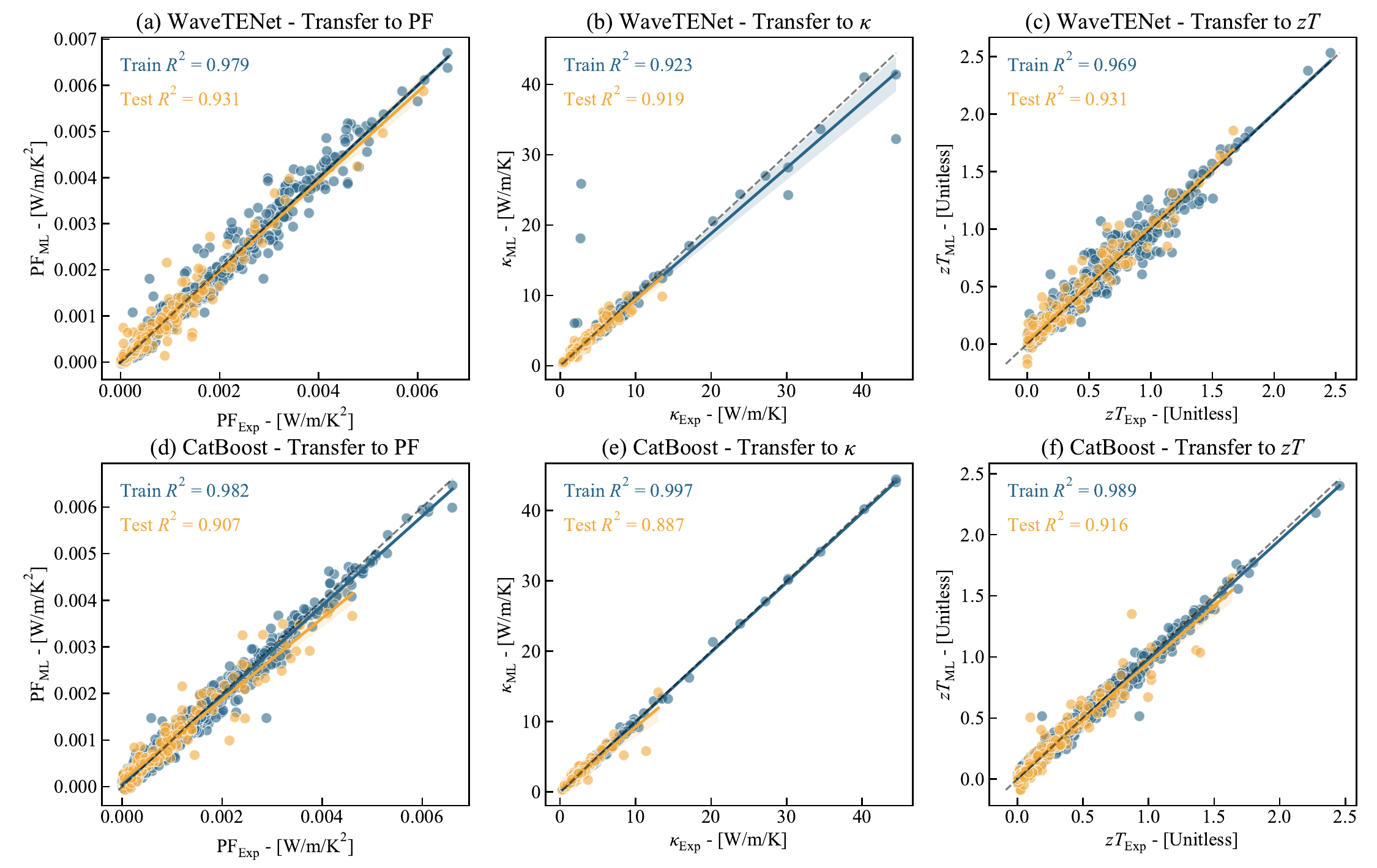}
    \caption{Performance comparison between WaveTENet and CatBoost on the transfer learning dataset. In the prediction tasks of PF, $\kappa$, and $zT$, WaveTENet consistently outperforms CatBoost in accuracy.}
    \label{fig:transfer}
\end{figure*}

To evaluate the performance of WaveTENet, we established a benchmark comprising DopNet, CatBoost~\cite{prokhorenkova2018catboost}, and a multilayer perceptron (MLP) model. The performance of DopNet was directly extracted from the original paper, while CatBoost and MLP were trained independently by us. We reported the average prediction performance measured over 10 repeated evaluations and ensured consistency in dataset partitioning for CatBoost and MLP. A comparative analysis of thermoelectric property predictions across different models is presented in Table~\ref{tab:benchmark}, where WaveTENet achieved the current \underline{S}tate \underline{o}f \underline{t}he \underline{A}rt (SOTA) performance.
\begin{figure*}
    \centering
    \includegraphics[scale=0.6]{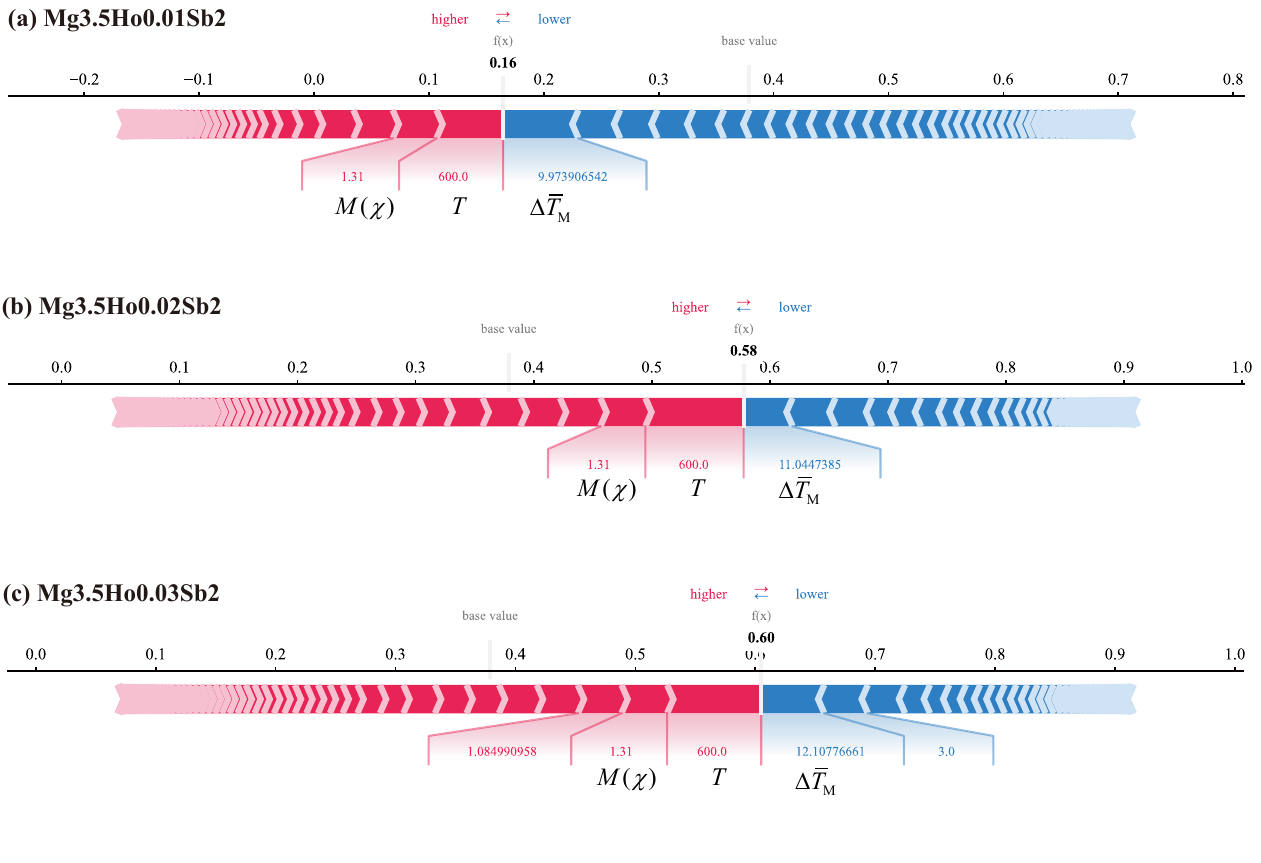}
    \caption{SHAP force plot for the Mg$_{3.5}$Ho$_{x}$Sb$_2$ system at 600~K. Bold numbers denote the surrogate model's estimated $zT$ values (not actual measurements). Numbers below the arrows represent feature values, with red indicating positive contributions and blue negative ones. Arrow length reflects feature influence, while $x$-axis tick marks quantify the impact magnitude. In subplot~(c), the unannotated values are attributed to intricate interactions among multiple features, rather than the effect of any individual descriptor alone.
    }
    \label{forceplot}
\end{figure*}
\begin{table}[h]
    \centering
    \caption{10-fold cross-validation results for each of the transport properties for the WaveTENet, DopNet, CatBoost and MLP models in terms of $R^2$. Bold values denotes the SOTA performance.}
    \begin{tabular}{lccccc}
        \toprule
        \textbf{Model} & $S$ & $\sigma$ & PF & $\kappa$ & $zT$ \\
        \midrule
        \textbf{WaveTENet}  & \textbf{0.977} & \textbf{0.984} & 0.981 & \textbf{0.972} & \textbf{0.963} \\
        DopNet     & 0.860 & 0.640 & 0.790 & 0.610 & 0.860 \\
        CatBoost   & 0.955 & 0.973 & \textbf{0.983} & 0.881 & 0.959 \\
        MLP        & 0.910 & 0.917 & 0.914 & 0.851 & 0.881 \\
        \botrule
    \end{tabular}
    \label{tab:benchmark}
\end{table}

The architecture of the MLP is $64 \times 64 \times 64$, with the number of training iterations determined by early stopping, where the tolerance is set to 30. The optimal hyperparameters for CatBoost were automatically tuned using Optuna~\cite{akiba2019optuna}, and the final hyperparameter configuration is provided in the Supplementary Information. Both MLP and CatBoost were trained exclusively on SIMD and Magpie descriptors without incorporating wavelet transforms. 
Interestingly, in predicting electronic transport properties, CatBoost and WaveTENet exhibited comparable predictive performance, with CatBoost even slightly surpassing WaveTENet in PF prediction. However, for $\kappa$, the inclusion of wavelet coefficients substantially enhanced accuracy. Generally, due to its tree-based structure, ensemble learning algorithms such as CatBoost tend to outperform deep learning models on non-smooth molecular feature datasets~\cite{treebetter,xia2023understanding,zeng2025ltc}. WaveTENet, by leveraging extensive residual stacking of linear layers, attains sufficient nonlinear learning capacity, enabling it to surpass conventional MLPs and achieve performance comparable to CatBoost in electronic transport predictions. Notably, WaveTENet significantly outperformed all other models in predicting $\kappa$, which may stem from its depth-driven nonlinear learning capacity or from the preservation of sample distinctiveness via identity mappings in linear layers and wavelet-derived features. A similar phenomenon has been observed in CraTENet's PF predictions~\cite{lantunes}.

To objectively evaluate the extrapolation capability of WaveTENet, we assess its performance on a new dataset using transfer learning and select CatBoost for comparison. Transfer learning refers to fine-tuning a model trained on a source task for a new task, thereby reducing training costs. In our study, the source task involves training a sufficiently well-performing predictive model on the ESTM dataset, which is then transferred to a new dataset to evaluate WaveTENet's performance across different datasets. The transfer learning dataset we employ originates from the UCSB dataset~\cite{ucsb,linxi}, with intrinsic materials manually removed. During training, we freeze the first five \texttt{deep residual blocks} of WaveTENet, leaving only the last one for parameter optimization.  

Notably, to eliminate the influence of feature vectors, we provide CatBoost with the same set of feature vectors, including those derived from wavelet transformations, i.e., $\mathcal{X}$ in Fig.~\ref{fig:net}(a). The results of transfer learning, as shown in Fig.~\ref{fig:transfer}, indicate that WaveTENet outperforms CatBoost in predicting PF, $\kappa$, and $zT$. Moreover, the pre-trained WaveTENet model demonstrates impressive performance on the new dataset, confirming its strong extrapolation capability and suitability for novel material discovery.

\subsection{Deepening Physical Interpretability via Sensitivity Analysis}
\label{physical-inter}
Among the three types of descriptors employed in our study, SIMD and wavelet coefficients lack explicit physical meaning, whereas Magpie provides statistical representations of the physical and chemical properties of constituent elements, making it inherently interpretable. As shown in Table~\ref{tab:benchmark}, the predictive accuracy of the CatBoost model based solely on Magpie descriptors remains within an acceptable range compared to WaveTENet for $zT$ prediction. Therefore, we select the CatBoost model with Magpie descriptors as a surrogate model for interpretability analysis.
\begin{figure*}
    \centering
    \includegraphics[scale=0.34]{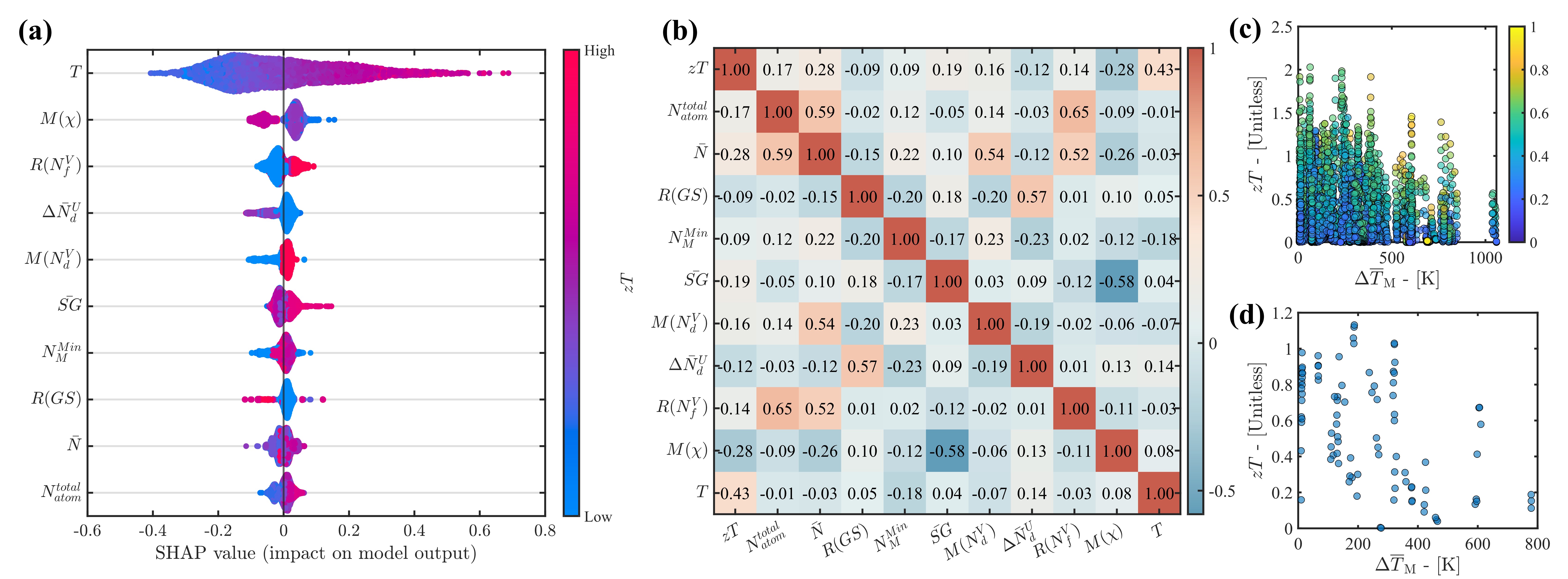}
    \caption{The figure provides an overview of feature importance and temperature-dependent behavior. (a) SHAP summary plot of the top 10 features ranked by the sum of absolute SHAP values. The vertical axis shows feature importance, while the horizontal axis indicates each feature’s impact on the model output. Each dot represents a sample, with vertical stacking showing density and color indicating feature values (red: high, blue: low). (b) Pearson correlation heatmap of the top 10 features, where cell values and color intensity represent the strength of correlation. (c) Scatter plot of $zT$ vs. $\Delta \bar{T}_{\rm M}$, with samples color-mapped by temperature $T$ to reveal thermal trends. (d) Same plot as (c), limited to samples at $T = 600\,\text{K}$ to highlight temperature-specific behavior.}
    \label{shapsum}
\end{figure*}

To uncover the potential relationships between elemental composition and $zT$, we employ SHAP~\cite{shap} to quantify the contribution of each descriptor to $zT$. The analysis focuses on the 10 key descriptors affecting $zT$, as illustrated in Fig.~\ref{shapsum}(a). Fig.~\ref{shapsum}(b) presents a heatmap of the Pearson correlation coefficients (PCC) among the key descriptors. Overall, the majority of descriptors exhibit negligible correlation, which serves as a prerequisite for applying SHAP, as kernel-based SHAP index does not account for dependencies among descriptors~\cite{shapdependence}. 

In Fig.~\ref{shapsum}(a), temperature $T$ emerges as the most significant feature influencing the $zT$ of thermoelectric materials, exhibiting a clear positive correlation. This is expected, as temperature is an explicit component in the definition of $zT$ in Eq.~(\ref{ztformula}). It is worth noting that the majority of the samples in our dataset correspond to temperatures ranging from 300~K to 700~K. In reality, when the temperature exceeds this range, the $zT$ value of some materials tends to decrease~\cite{zttemplow}. However, if we confine our analysis to the dataset used in this study, this conclusion remains valid, as it is well established that for most thermoelectric materials, $zT$ generally increases with temperature within the 300~K to 700~K range.  

Beyond the direct mathematical dependence of $zT$ on $T$, temperature also indirectly influences $zT$ by affecting $\sigma$, $S$, and $\kappa$. The electrical conductivity $\sigma$ and the Seebeck coefficient $S$ are defined as:
\begin{align}
    \label{sigmadefine} \sigma&=nq\mu\\
    \label{sdefine} S &= \frac{\pi^2 k_B^2 T}{3e} \left[ \frac{1}{\sigma} \frac{\text{d}\sigma(E)}{\text{d}E} \right]_{E=E_{\rm F}}
\end{align}
where $n$ denotes the carrier concentration, representing the number of free electrons or holes per unit volume, $q$ is the elementary charge, and $\mu$ is the carrier mobility, which quantifies the drift velocity of carriers under an applied electric field. Hence, electrical conductivity is jointly determined by carrier concentration and mobility. Eq.~(\ref{sdefine}), known as the Mott relation~\cite{mott}, describes the relationship between the Seebeck coefficient $S$ and the energy-dependent conductivity gradient. The prefactor $\frac{\pi^2 k_B^2 T}{3e}$ is a temperature-dependent coefficient, where $k_B$ is the Boltzmann constant, $T$ is the absolute temperature, and $e$ is the elementary charge. The term inside the brackets, $\left[ \frac{\text{d}\sigma(E)}{\text{d}E} \right]_{E=E_F}$, represents the normalized energy derivative of the conductivity, evaluated at the Fermi level $E_F$. This quantity characterizes the asymmetry of carrier transport properties. 
An increase in temperature generally leads to a decline in carrier mobility~\cite{qianyilv}, primarily due to enhanced carrier scattering~\cite{sanshe}. According to Eq.~(\ref{sigmadefine}) and Eq.~(\ref{sdefine}), the reduction in carrier mobility directly results in a decrease in $\sigma$, whereas both the decrease in $\sigma$ and the increase in temperature contribute to an increase in $S$. The temperature dependence of $S$ and $\sigma$ exhibits a competing behavior; however, within a given temperature range, a peak in the PF is typically observed~\cite{pffengzhi}. A similar trend is also commonly found in their variations with carrier concentration.

Beyond electrical transport, the temperature dependence of thermal conductivity must also be considered. In thermoelectric semiconductors, lattice thermal conductivity $\kappa_{\rm L}$ generally dominates over electronic thermal conductivity $\kappa_{\rm e}$ in magnitude and serves as the primary factor. At intermediate to high temperatures, Umklapp scattering intensifies, leading to a reduction in the phonon mean free path and consequently lowering $\kappa_{\rm L}$~\cite{holland1966thermal}. In summary, $zT$ exhibits an increasing trend with temperature. However, it is important to note that this conclusion is only valid within the approximate temperature range of 300~K to 700~K.

$M(\chi)$ represents the mode of the electronegativity of elements in a material, defined as:
\begin{align}  
    M(\chi) = \arg\max_{\chi} f(\chi)  
\end{align}  
where $\chi$ denotes the electronegativity of an element, and $f(\chi)$ represents the frequency function of electronegativity occurrences in the chemical formula. In simple terms, $M(\chi)$ reflects the electronegativity of the dominant element in the composition.
% \begin{figure}[h]
%     \centering
%     \includegraphics[scale=0.3]{zt-tm.jpg}
%     \caption{Dependence of $zT$ on $\Delta \bar T_{\rm M}$. (a) Each sample is color-mapped according to its corresponding $T$; (b) Only samples with $T=600~K$ are selected.}
%     \label{fig:avg-dev}
% \end{figure}

Electronegativity essentially describes an element’s ability to attract electrons when forming chemical bonds~\cite{electronegativitydefine}. Elements with high electronegativity tend to bind electrons more strongly, thereby reducing the free carrier concentration and consequently decreasing electrical conductivity. By doping elements with lower electronegativity, the covalency of chemical bonds can be enhanced, which weakens carrier-phonon coupling, leading to a lower effective mass and thus improving carrier mobility~\cite{ren2017enhancing}.  

Although, as shown in Eq.~(\ref{sdefine}), a decrease in $\sigma$ may enhance $S$, electronegativity also plays a role in modulating the Fermi level. A higher electronegativity of the dominant element suggests that doping enhances the p-type behavior of the material, causing the Fermi level to shift downward (closer to the valence band). This results in a slower variation of the electronic density of states, leading to a decrease in the term $\left[ \frac{\text{d}\sigma(E)}{\text{d}E} \right]_{E=E_F}$, which in turn negatively impacts the Seebeck coefficients~\cite{doss,chen2012effect}.

Fig.~\ref{shapsum} presents the global interpretability analysis based on SHAP, while Fig.~\ref{forceplot} provides a local explanation for individual samples. Taking the Mg$_{3.5}$Ho$_{x}$Sb$_2$ system mentioned in Sec.~\ref{nonlinear} as an example, the $zT$ value of this system is primarily positively influenced by $T$ and $M(\chi)$, whereas the deviation of the average melting point of the constituent elements, $\Delta \bar T_{\rm M}$, exerts the most significant negative effect. 

As shown in Fig.~\ref{forceplot}, the dominant factor affecting Mg$_{3.5}$Ho$_{x}$Sb$_2$ is the fluctuation in the melting point caused by variations in Ho doping concentration. Specifically, increasing the Ho fraction by 0.01 leads to an approximately 10\% increase in $\Delta \bar T_{\rm M}$. As depicted in Fig.~\ref{shapsum}(d), under the constraint of $T=600{\rm K}$, the upper limit of $zT$ appears to decrease as $\Delta \bar T_{\rm M}$ increases. 

However, the mechanism by which $\Delta \bar T_{\rm M}$ influences $zT$ remains elusive. From a global perspective (Fig.~\ref{shapsum}(c)), $zT$ does not exhibit a consistent trend with $\Delta \bar T_{\rm M}$. There exist numerous low-$zT$ materials even when $\Delta \bar T_{\rm M}$ is low, while high-$zT$ materials are often associated with elevated $T$, showing no apparent correlation with $\Delta \bar T_{\rm M}$. 
While $\Delta \bar T_{\rm M}$ can be considered the dominant factor in the Mg$_{3.5}$Ho$_{x}$Sb$_2$ system, its generalization to a broader scope is not justifiable.

\subsection{Multi-Objective Optimization using NSGA-III}

% 不同于传统的Genetic Algorithms，Non-dominated Sorting Genetic Algorithm III (NSGA-III)专门适用于三个或更多目标的协同优化问题。In our case, 为了探索具有高$zT$值的新型热电材料，需要同时优化材料的元素和组分，使之具有尽可能高的$|S|$和$\sigma$、尽可能低的$\kappa$： 
% \begin{align}
%     \max_{\mathcal{X}} \left( |S(\mathcal{X})|, \sigma(\mathcal{X}), \frac{1}{\kappa(\mathcal{X})} \right)
% \end{align}
% 出于问题简化，我们将$\min_{\mathcal{X}} \kappa(\mathcal{X})$等价变换为$\max_{\mathcal{X}} \frac{1}{\kappa(\mathcal{X})}$. 
% 由于元素、分数所构成的集合是难以穷举的，因此需要用优化算法尽可能高效的搜索材料空间。NSGA-III模拟自然选择和遗传学原理，通过模拟生物进化过程来求解问题，其中“个体”通过遗传、变异和选择等过程逐渐提升适应度，最终找到最优解。我们将遗传学概念迁移到材料设计中，一种材料就是一个生物个体，将组成元素及分数视作基因，将材料TE性能$|S|$,$\sigma$和$\frac{1}{\kappa}$视作个体的适应度。
\begin{figure*}
    \centering
    \includegraphics[scale=0.45]{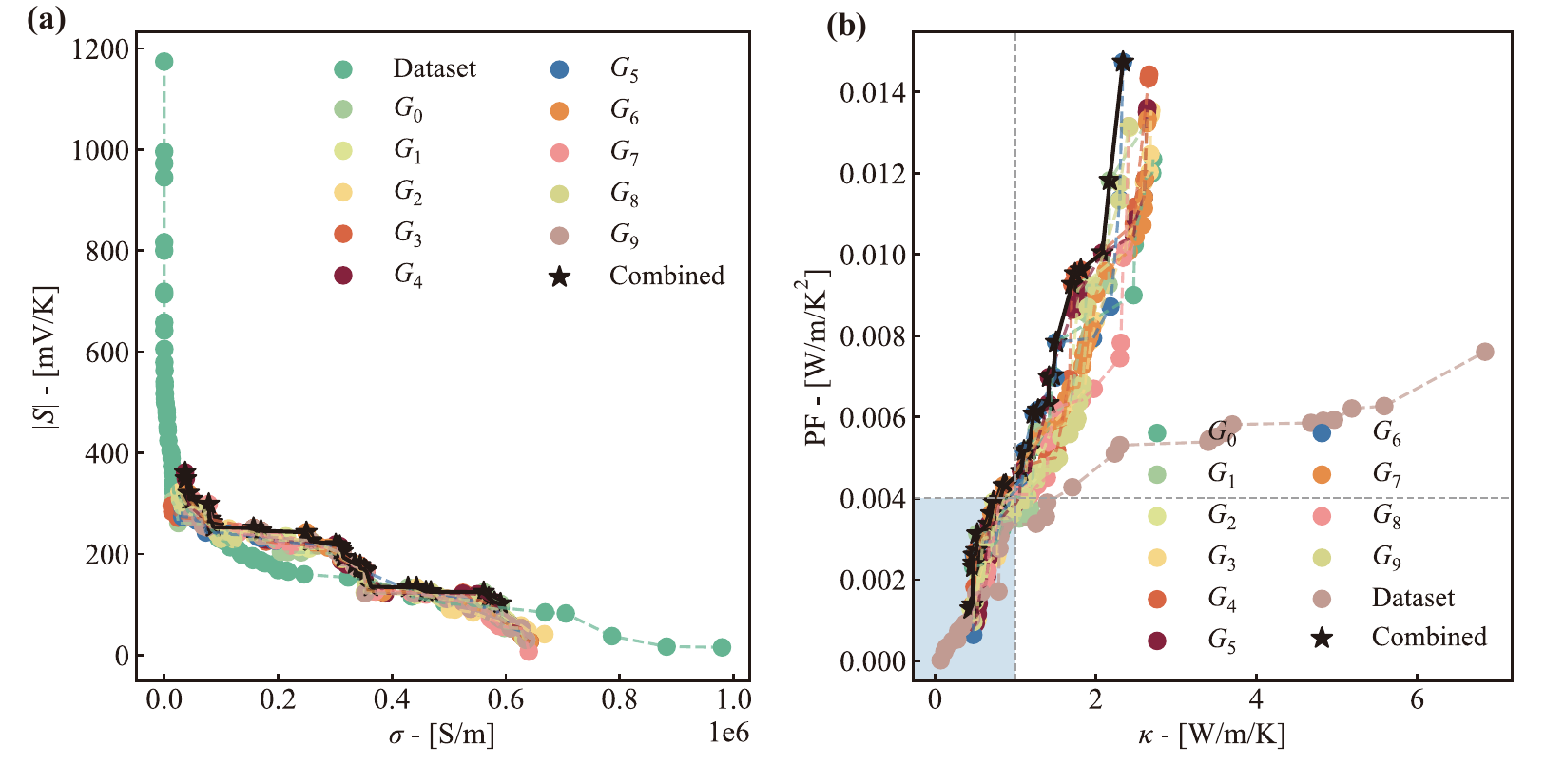}
    \caption{Comparison of the Pareto-fronts before and after MOO. Different colors of circular scatter points and connecting lines represent the optimal candidates and corresponding Pareto-fronts from the initial dataset and different generations, while the black pentagram symbolizes the optimal results after merging across all generations. (a) Focuses on $S$ and $\sigma$, considering them as a unified optimization direction, with candidates selected based solely on the absolute value of $S$. (b) Focuses on PF and $\kappa$, where the candidates in the lower-left region of the gray dashed line more closely align with the Pareto-fronts of the initial dataset.
   }
    \label{fig:pareto}
\end{figure*}
\begin{table*}
    \centering
    \caption{Pareto-optimal candidate materials after screening.}
    \begin{tabular}{l l c c c c c c c}
        \toprule
        \textbf{Base} & \textbf{Formula} & \textbf{Gen.} & \textbf{Temperature} & $S_{\rm pred}$ & $\sigma_{\rm pred}$ & PF & $\kappa_{\rm pred}$ & $zT$ \\
        \midrule
        BiSe & BiSeGe$_{0.12}$Bi$_{0.37}$Y$_{0.16}$ & 2 & 600 & -200.89 & 31664.66 & 0.00128 & 0.4427 & 1.7319 \\
        BiSe & BiSeBi$_{0.25}$Y$_{0.18}$ & 7 & 600 & -237.27 & 24423.96 & 0.00137 & 0.4644 & 1.7764 \\
        BiSe & BiSeAg$_{0.32}$Bi$_{0.39}$ & 0 & 600 & -254.01 & 35734.53 & 0.00231 & 0.4759 & 2.9070 \\
        BiSe & \underline{BiSeSb$_{0.2}$Y$_{0.13}$Se$_{0.33}$} & 1 & 367 & -183.36 & 71620.65 & 0.00241 & 0.4806 & 1.8373 \\
        BiSe & BiSeAg$_{0.37}$Bi$_{0.46}$ & 6 & 600 & -266.28 & 36687.65 & 0.00260 & 0.4839 & 3.2258 \\
        BiSe & BiSeGe$_{0.06}$Ag$_{0.36}$Bi$_{0.39}$ & 7 & 600 & -247.77 & 42446.70 & 0.00261 & 0.4990 & 3.1332 \\
        BiSe & BiSeZn$_{0.26}$Bi$_{0.50}$Ag$_{0.11}$ & 4 & 600 & -263.10 & 39256.36 & 0.00272 & 0.5156 & 3.1620 \\
        Bi$_2$Te$_3$ & Bi$_2$Te$_3$Y$_{0.23}$Ag$_{0.36}$ & 1 & 433 & -228.76 & 59676.69 & 0.00312 & 0.5265 & 2.5701 \\
        Bi$_2$Te$_3$ & Bi$_2$Te$_3$Y$_{0.44}$Sn$_{0.23}$ & 5 & 600 & -203.63 & 81164.93 & 0.00337 & 0.6710 & 3.0094 \\
        Bi$_2$Te$_3$ & Bi$_2$Te$_3$Sn$_{0.19}$Y$_{0.50}$ & 7 & 600 & -210.32 & 81858.41 & 0.00362 & 0.7015 & 3.0969 \\
        Bi$_2$Te$_3$ & Bi$_2$Te$_3$Se$_{0.18}$Y$_{0.18}$Sb$_{0.17}$ & 2 & 433 & -219.34 & 81344.72 & 0.00391 & 0.7267 & 2.3335 \\
        Bi$_2$Te$_3$ & Bi$_2$Te$_3$Zn$_{0.41}$Y$_{0.16}$Sb$_{0.25}$ & 4 & 433 & -206.58 & 93000.80 & 0.00397 & 0.8109 & 2.1209 \\
        Sb$_2$Te$_3$ & Sb$_2$Te$_3$Te$_{0.43}$Y$_{0.18}$Bi$_{0.39}$ & 7 & 300 & 233.58 & 79066.91 & 0.00431 & 0.8552 & 1.5132 \\
        Sb$_2$Te$_3$ & Sb$_2$Te$_3$Te$_{0.43}$Y$_{0.18}$Bi$_{0.39}$ & 7 & 333 & 232.67 & 79952.61 & 0.00433 & 0.8796 & 1.6403 \\
        \botrule
    \end{tabular}
    \label{tab:thermoelectric-results}
\end{table*}

Unlike traditional Genetic Algorithms (GA)~\cite{Holland2012GeneticA}, the NSGA-III~\cite{nsgaiii} is specifically designed for MOO involving three or more objectives. In our case, to discover potential thermoelectric materials, the algorithm should be capable of simultaneously optimizing both the elemental composition and fraction to maximize $|S|$ and $\sigma$, while minimizing $\kappa$:
\begin{align}
    \max_{\mathcal{X}} \left( |S(\mathcal{X})|, \sigma(\mathcal{X}), \frac{1}{\kappa(\mathcal{X})} \right)
\end{align}
For simplicity, we transform $\min_{\mathcal{X}} \kappa(\mathcal{X})$ into the equivalent maximization problem $\max_{\mathcal{X}} \frac{1}{\kappa(\mathcal{X})}$. Although it is feasible to directly optimize $zT$ as a single objective using traditional GA or Bayesian optimization, the antagonistic nature between $S$ and $\sigma$ (see Sec.~\ref{physical-inter}) makes it impractical to achieve simultaneously high $S$ and $\sigma$ leading to an exceptionally high $zT$. Therefore, MOO is more advantageous as it ensures the reasonableness of newly generated materials while also tracing the origins of high $zT$.
\begin{table}[h]
    \centering
    \renewcommand{\arraystretch}{1.2}
    \caption{Multi-objective optimization (MOO) parameters.}
    \begin{tabular}{lc}
        \toprule
        \textbf{MOO Parameters} & \textbf{Value} \\
        \midrule
        Population size & $N=300$ \\
        Generations & $T=10$ \\
        Crossover rate & $p_{\rm c}=0.9$ \\
        Polynomial mutation & $\eta=20$ \\
        Reference direction & $H=12$ \\
        Sampling & LHS \\
        Base materials & Bi$_2$Te$_3$, Sb$_2$Te$_3$, PbTe, SnTe, GeTe, \\
                      & SiGe, Mg$_2$Si, BiSe \\
        Dopants & Bi, Sb, Te, Se, Sn, Pb, Ag, Cu, Ge, \\ & Zn, Y \\
        Temperature & $[300, 600]$ \\
        \botrule
    \end{tabular}
    \label{tab:MOO_parameters}
\end{table}

Since the set of elements and their fractions is too large to exhaustively enumerate, optimization algorithms are necessary to efficiently search the material space. NSGA-III mimics natural selection and principles of genetics, solving the problem through the simulation of biological evolution, where ``individuals'' gradually improve their fitness via genetic processes such as inheritance, mutation, and selection, ultimately finding the optimal solution~\cite{nsgaiii}. We transfer these genetic concepts to material design, treating a compound as an individual organism, with the elemental composition and fractions as its genes, and the material's thermoelectric properties (e.g., $|S|$, $\sigma$, and $\frac{1}{\kappa}$) as its fitness.

To identify high-performance thermoelectric materials suitable for wearable thermoelectric devices, the temperature domain is restricted to the range of 300~K to 310~K. In the generation of virtual materials, we select commonly used base materials and introduce dopants at compositions ranging from 0.05 to 1.0, with an increment of 0.05. Candidates of base materials and dopants are summarized in Table~\ref{tab:MOO_parameters}.

The NSGA-III MOO framework is implemented using the Python library \texttt{pymoo-0.6.1.3}~\cite{pymoo}. For the parameter configuration of NSGA-III, Latin Hypercube Sampling (LHS)~\cite{lhs} is employed to ensure a uniform distribution of the initial population within the variable space, thereby enhancing diversity. The population size per generation, $T$, is set to 50, meaning that each generation produces 300 candidate solutions. A larger population facilitates broader exploration of the material space but incurs higher computational costs, whereas a smaller population converges faster but risks being trapped in local optima~\cite{tanabe2017impact}. The crossover probability ($p_{\rm c}$) is set to 0.9, indicating that 90\% of individuals undergo crossover while 10\% remain unchanged—an emulation of natural reproduction processes~\cite{yi2020behavior}. To prevent premature convergence, polynomial mutation (PM) is incorporated to introduce random variations, mimicking genetic mutations in natural evolution~\cite{rosenthal2014impact}.

Fig.~\ref{fig:pareto} illustrates the Pareto-front of each generation throughout the NSGA-III optimization process, with the optimal virtual candidate set, post-merging, clearly annotated. In Fig.~\ref{fig:pareto}(a), the distributions of $|S|$ and $\sigma$ for the virtual candidates generally align with those of the original dataset. However, in Fig.~\ref{fig:pareto}(b), the corresponding distributions of PF and $\kappa$ deviate significantly from the real data. Specifically, some candidates demonstrate unexpectedly elevated PF values, while all candidates show $\kappa$ values below 2.5 W/m/K. These excessively high PF values arise from error propagation, where minor overestimations in $S$ and $\sigma$ are amplified through their multiplicative relationship. In the range of $\sigma \in [0.15 \times 10^6, 0.4 \times 10^6]$, several samples' objectives do not closely match the Pareto-front of the original dataset. Although the individual predictions for $\sigma$ and $S$ are reliable, the error in $PF = S^2 \sigma$ is substantially magnified. As observed during surrogate model training (see Fig.~\ref{fig:performance}) and transfer learning (see Fig.~\ref{fig:transfer}), the model’s accuracy for $\kappa$ estimation is the lowest. Consequently, in Fig.~\ref{fig:pareto}(b), only the candidates in the lower-left region of the gray dashed line are considered to have validation value. The remaining samples, which display unrealistically high PF values and relatively low $\kappa$, significantly diverge from the distribution of the original dataset and, therefore, do not hold reference value.
\begin{figure}[h]
    \centering
    \includegraphics[scale=0.47]{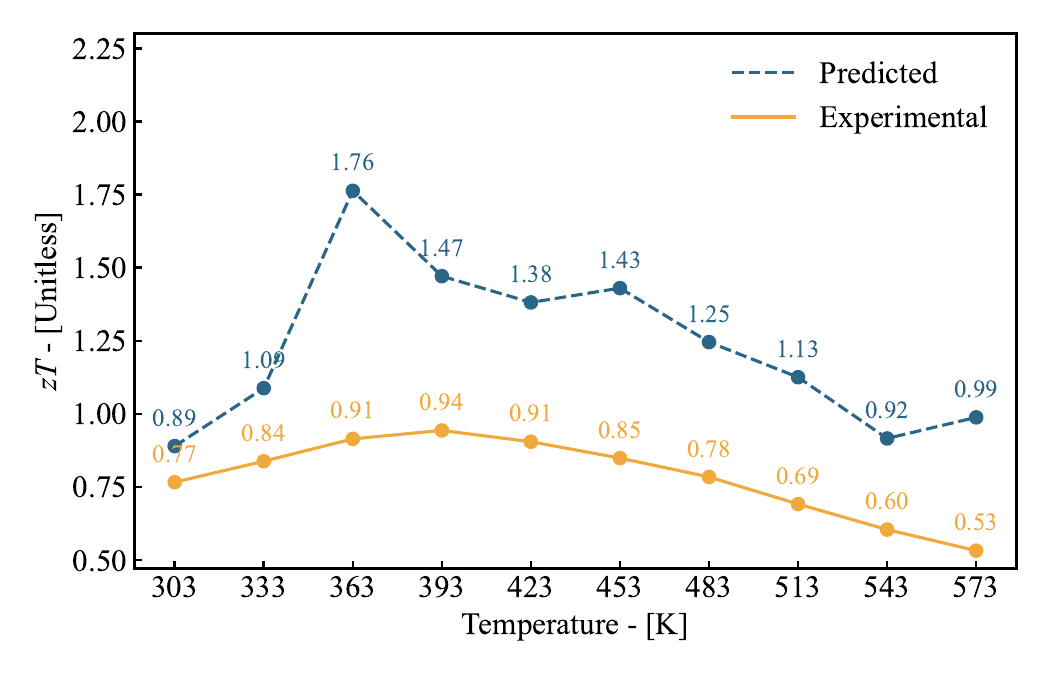}
    \caption{Comparison of the predicted and measured $zT$ values for BiSeSb$_{0.2}$Y$_{0.13}$Se$_{0.33}$ (Bi$_{0.75}$Sb$_{0.15}$Y$_{0.10}$Se). The predicted $zT$ is obtained from the WaveTENet model, while the measured value is derived from experimental data.}
    \label{fig:compare}
\end{figure}

Table~\ref{tab:thermoelectric-results} summarizes the screened Pareto-optimal candidate materials, all exhibiting PF values above 0.001~W/m/K$^2$, $\kappa$ below 1.0~W/m/K, and $zT$ exceeding 1.5, with most surpassing 2.0. Despite promising predictions, model uncertainty may lead to deviations in actual performance, necessitating experimental validation.
The candidates are primarily derived from BiSe, Bi$_2$Te$_3$, and Sb$_2$Te$_3$, with only Sb$_2$Te$_3$-based compounds being $p$-type. To differentiate hosts from dopants, identical elements in the formulas were not merged. Considering material cost, Te-containing candidates were excluded. Among the BiSe-based materials, BiSeSb${0.2}$Y${0.13}$Se$_{0.33}$, which achieves a peak $zT$ at 367~K, is selected for experimental validation due to its excellent performance without relying on high-temperature operation.

Here, the formula ``BiSeSb${0.2}$Y${0.13}$Se$_{0.33}$'' was equivalently replaced with ``Bi$_{0.75}$Sb$_{0.15}$Y$_{0.10}$Se'' for easier synthesis. As shown in Fig.~\ref{fig:compare}, the measured $zT$ of Bi$_{0.75}$Sb$_{0.15}$Y$_{0.10}$Se at $T = 363$~K was 0.91, which is much lower than the predicted value of 1.76. However, this material still demonstrates excellent thermoelectric performance, with a $zT$ at around 360~K that surpasses recent $n$-type materials such as Ag$_8$SiSe$_6$ ($\sim$0.65 at 360~K)~\cite{wang2024new}, Pb$_{1.006}$S$_{0.6}$Se$_{0.4}$ ($<0.5$ at 360~K)~\cite{wang2024realizing}, and Mg$_3$SbBi ($<0.8$ at 360~K)~\cite{zuo2025atomic}. In addition to its high $zT$ value, Bi$_{0.75}$Sb$_{0.15}$Y$_{0.10}$Se also features a relatively simple synthesis process (see Sec.~\ref{sec:zhibei}), along with advantages such as low cost and environmental friendliness, making it a promising candidate for practical applications.

Although the deep learning model is subject to uncertainty and has overestimated the $zT$ value of Bi$_{0.75}$Sb$_{0.15}$Y$_{0.10}$Se, it has accurately captured the temperature dependence of its thermoelectric performance, namely, the trend of $zT$ increasing and then decreasing over the range of 303–573~K. The prediction uncertainty arises not only from the inherent limitations of the model itself but also from the lack of representation of synthesis conditions. Even with identical chemical compositions, different synthesis methods can lead to nonnegligible variations in material performance~\cite{wang2025process, abdellahi2015effect, sotelo2015effect}.

\section{Discussion}

This work demonstrates a deep learning model called WaveTENet, which innovatively leverages wavelet transforms for feature enhancement, enabling it to accurately capture both local and global characteristics of materials. This approach significantly mitigates the interference caused by the nonlinear effects between material composition and performance on predictions. To the best of our knowledge, WaveTENet achieves state-of-the-art performance in predicting the thermoelectric properties of doped materials. Furthermore, through sensitivity analysis, we have investigated the impact of the electronegativity and melting point deviation of the constituent elements on thermoelectric performance, enhancing the model's physical interpretability. Considering the potential antagonistic relationship between key properties, we select NSGA-III as the optimization algorithm, performing multi-objective collaborative optimization on $S$, $\sigma$, and $\kappa$, which ultimately identifies a series of promising candidates with excellent thermoelectric performance. Considering preparation difficulty and application value, we chose Bi$_{0.75}$Sb$_{0.15}$Y$_{0.10}$Se for experimental validation. Although its actual $zT$ values is lower than predicted, it still demonstrates superior thermoelectric performance, surpassing many commercial $n$-type thermoelectric materials. 

Certainly, the predictions made by WaveTENet inevitably involve some degree of uncertainty. On one hand, achieving perfect predictions through deep learning remains impractical at present; on the other hand, the existing data lacks characterization of material fabrication processes, leading us to make necessary simplifications in this work. However, as of now, WaveTENet still represents an efficient and relatively accurate method for predicting the thermoelectric properties of doped materials, and the surrogate model-based MOO framework also provides valuable insights for materials inverse design.

\section{Methods}

\subsection{System-Identified Material Descriptor}
\label{desc-method}

As presented in Eq.~(\ref{simdeq}), the SIMD vector consists of $\mathbf{x}_{s}$, $\mathbf{c}_{s}$, $\mathbf{w}_{s}^{(K)}$, and $\mathbf{w}_{s}^{(K)}$. Specifically, $\mathbf{x}_{s}$ encodes the elemental fractions in the material composition using a sparse representation:  
\begin{align}
x_{i}= 
\begin{cases}
r_{e}, & \text{if } i=n_{e} \text{ and } e \in s, \\  
0, & \text{otherwise}.  
\end{cases}
\end{align}  
in this formula, $e$ denotes an element in the composition $s$, $r_e$ represents its fraction, and $n_e$ corresponds to its atomic number. The SIMD framework~\cite{na2022public} accounts for 100 elements, spanning from H to Fm. The vector $\mathbf{c}_{s}$ represents the conditional vector, which, in principle, can incorporate information related to material synthesis, such as temperature, pressure, and cooling time. However, the ESTM and UCSB datasets provide only temperature information, making $\mathbf{c}_{s}$ an $n \times 1$ vector. Together, $\mathbf{x}_{s}$ and $\mathbf{c}_{s}$ determine the system vector $\mathbf{w}_{s}^{(K)}$:  
\begin{align}
    \begin{bmatrix}
        x_{11} & \cdots & x_{1M} & c_{11} & \cdots & c_{1L} \\  
        \vdots & \ddots & \vdots & \vdots & \ddots & \vdots \\  
        x_{|u|1} & \cdots & x_{|u|M} & c_{|u|1} & \cdots & c_{|u|L}  
    \end{bmatrix}  
    \begin{bmatrix}
        w_{1} \\  
        \vdots \\  
        w_{d}  
    \end{bmatrix}  
    =
    \begin{bmatrix}
        y_{1} \\  
        \vdots \\  
        y_{|u|}
    \end{bmatrix}
    \label{matrix}
\end{align}  
here $|u|$ denotes the number of materials in cluster $u$, and $\mathcal{y}$ represents the target property. The solution of Eq.~(\ref{matrix}), $\mathbf{w}=[w_1,\dots,w_d]^{\rm T}$, has a dimensionality of $d=M+L$, corresponding to the combined dimensions of $\mathbf{x}_{s}$ and $\mathbf{c}_{s}$. Notably, $\mathbf{w}$ represents only the system vector of a single cluster, while $\mathbf{w}_{s}^{(K)}$ further incorporates information from the $K$ nearest clusters to enhance representation:  
\begin{align}
    {\mathbf{w}}_{s}^{(K)}=\sum_{u \in \mathcal{N}_{s}} \frac{\frac{1}{r(a_u,a_s)}}{\sum_{h \in \mathcal{N}_{s}}\frac{1}{r(a_h,a_s)}} {\mathbf{w}}^{(u)}
    \label{system-vec}
\end{align}  
where $\mathcal{N}_{s}$ denotes the set of $K$ nearest material clusters to the current cluster. The vector $a_h$ represents the anchor point of a material cluster in the anchor space, while $a_s$ corresponds to the anchor point of the input composition $s$. The influence of samples from different clusters on the current cluster $u$ is quantified using a distance function $r(\cdot)$, which measures geometric proximity. 

The target statistical vector $\mathbf{v}$ comprises the mean, standard deviation, minimum, and maximum of the target properties within a material cluster. It is formally defined as:
\begin{align}
    v = \left[\overline{y}, y_{\sigma}, y_{\rm min}, y_{\rm max}\right]
\end{align}
where $\overline{y}=\frac{1}{|u|}\sum_{i = 1}^{|u|}y_{i}$ represents the mean target property, $y_{\sigma}=\sqrt{\frac{\sum_{i = 1}^{|u|}(y_{i}-\overline{y})^{2}}{|u|}}$ denotes the standard deviation, and $y_{\rm min}$ and $y_{\rm max}$ correspond to the minimum and maximum values within the target set $\{y_{1}, y_{2}, \ldots, y_{|u|}\}$, respectively. Similar to Eq.~(\ref{system-vec}), the target statistical vector is also processed in the anchor space:
\begin{align}
    {\mathbf{v}}_{s}^{(K)}=\sum_{u \in \mathcal{N}_{s}} \frac{\frac{1}{r(a_u,a_s)}}{\sum_{h \in \mathcal{N}_{s}}\frac{1}{r(a_h,a_s)}} {\mathbf{v}}^{(u)}
\end{align}  

Ultimately, the SIMD-generated vector conforms to the structure of Eq.~(\ref{simdeq}).

The Materials-Agnostic Platform for Informatics and Exploration (Magpie)~\cite{ward2016general} descriptor enables the quantitative identification of physical and chemical properties based solely on elemental composition, without considering material structure. These properties include, but are not limited to, stoichiometric attributes, elemental properties, and ionization characteristics. The specific feature types used in this study are detailed in the Supplementary Information.

\subsection{Discrete Wavelet Transformation}
The DWT is a widely used technique for multi-resolution analysis, enabling the decomposition of signals into localized time-frequency components. Unlike Fourier transform, which provides only frequency-domain representation, DWT retains both time and frequency information, making it well-suited for feature extraction in structured data. In this study, we employ the Haar wavelet~\cite{lepik2014haar}, the simplest and most computationally efficient wavelet, to enhance feature representation.

DWT decomposes an input signal $x[n]$ into two sets of coefficients at each level: 
\begin{itemize}
    \item The approximation coefficients $a_j$, which capture low-frequency components and preserve the overall trend, and
    \item The detail coefficients $d_j$, which highlight high-frequency variations and fine details. The transformation is performed using a low-pass filter $h[n]$ and a high-pass filter $g[n]$, followed by a downsampling operation:
\end{itemize}

The defination of DWT can be expressed as: 
\begin{align}
a_j[k] &= \sum_{n} x[n] h[2k - n]\\
d_j[k] &= \sum_{n} x[n] g[2k - n]
\end{align}

For the Haar wavelet, the filters are defined as:
\begin{align}
h[n] &= \frac{1}{\sqrt{2}} [1, 1]\\ 
g[n] &= \frac{1}{\sqrt{2}} [1, -1]
\end{align}
which ensures efficient decomposition while preserving key structural information. 

In our approach, each feature undergoes a single-level Haar wavelet decomposition. The resulting approximation and detail coefficients are concatenated with the original feature set to enrich the representation. This augmentation process effectively increases feature diversity while maintaining computational efficiency. The choice of a single-level decomposition is based on empirical validation, ensuring that essential information is retained without introducing excessive redundancy.

In our work, the choice of DWT over the Fast Fourier Transform (FFT)~\cite{rao2011fast}, which is also suitable for discrete signals, is due to DWT's superior capability in detecting transient signal changes~\cite{robertson1996wavelets}. Both SIMD and Magpie descriptors exhibit sparsity, particularly in the element fraction encoding of SIMD and the one-hot vector component of Magpie. Furthermore, the differences between materials in the same doped system are subtle in terms of their descriptors, and DWT is better suited for capturing high-frequency variations than FFT~\cite{deokar2014integrated}.

\subsection{SHAP Index}
The SHAP values are computed following the method proposed by Lundberg and Lee~\cite{shap} and implemented using the \texttt{shap} package in Python. SHAP, derived from game theory's Shapley values, is a model interpretability approach that quantifies each player's contribution to the overall outcome. In this work we leverage SHAP values to assess the contribution of individual Magpie descriptors $\mathcal{M} \in \mathcal{N} = \{1,2,\cdots,N\}$ to the predicted $zT$ value. For this purpose, the ``players'' are abstracted as different descriptor dimensions, while the ``outcome'' corresponds to the $zT$ value. The Shapley value is computed as follows:
\begin{align}
    \phi_j = \sum_{\mathcal{M}\subseteq\mathcal{N} \setminus \left\{j\right\}}&
    \frac{\left|\mathcal{M}\right|! \left(N - \left|\mathcal{M}\right| - 1\right)!}{N!} 
    \left[v\left(\mathcal{M}\cup\left\{j\right\} \right) - v\left(\mathcal{M}\right)\right] \notag \\
    & j=1,\dots,N.
\end{align}
where $|\mathcal{M}|$ represents the dimensionality of the feature vector~\cite{aas2021explaining}. $ v(\mathcal{M}) $ represents the contribution value, i.e., the expected model output given a specific feature subset $\mathcal{M}$. It quantifies the extent to which the subset $\mathcal{M}$ influences the model's prediction. Given a surrogate model $zT(\mathcal{X})$, $ v(\mathcal{M}) $ can be formally expressed as:  
\begin{align}  
    &zT(\mathcal{X}^*)=\phi_0+\sum_{j=1}^N \phi_j^*\\
    &v(\mathcal{M}) = E\left[zT(\mathcal{X}) \mid \mathcal{X}_\mathcal{M} = \mathcal{X}_\mathcal{M}^*\right]  
\end{align}
in the above equations, $\phi_0$ represents the mean Shapley value of the surrogate model $zT(\mathcal{X})$, which is also the model's average prediction. The term $\mathcal{X}^*$ refers to a specific selected feature subset, and $\phi_j^*$ denotes the Shapley value of the $j$th feature when $\mathcal{X} = \mathcal{X}^*$. This process is more intuitively illustrated in Fig.~\ref{forceplot}, where the ``base\_value'' corresponds to the mean predicted $zT$ value of the surrogate model, i.e., $\phi_0$. Under the interplay of multiple features such as $M(\chi)$ and $T$, each with their respective Shapley values $\bm \phi_j$, the model's specific prediction is determined~\cite{purcell2023accelerating}.

\subsection{Synthesis methodology of Bi$_{0.75}$Sb$_{0.15}$Y$_{0.10}$Se}
\label{sec:zhibei}

The Bi$_{0.75}$Sb$_{0.15}$Y$_{0.10}$Se compound was synthesized using bismuth (Bi, Afla Aesar, 99.99\%, granules), antimony (Sb, Afla Aesar, 99.99\%, powder), yttrium (Y, Afla Aesar, 99.99\%, powder), and selenium (Se, Afla Aesar, 99.99\%, powder) as raw materials. These compounds were prepared by high-energy ball milling for 10 hours in a stainless steel ball mill jar under a pure argon atmosphere. The synthesized powder samples were examined using an X-ray diffractometer (XRD, Rigaku Corporation, Japan), and the peak positions of the samples were compared to standard cards using the Jade 6 program.

\subsection{Pseudocode of NSGA-III}
\label{nsga}
\begin{algorithm}[H]
    \caption{NSGA-III for Multi-objective thermoelectric Material Optimization}
    \label{alg:nsga3}
    \KwIn{Population size $N=300$, generations $T=10$, crossover rate $p_{\rm c}=0.9$, polynomial mutation PM $(\eta=20)$, reference directions $H=12$}
    \KwOut{Pareto-front $\mathcal{P}^*$, optimal composition $\mathcal{X}^*$}

    \textbf{Step 1: Initialization}
    \begin{itemize}
        \item Generate $H$ reference directions $\bm{w}_1, ..., \bm{w}_H$\;
        \item Initialize population $P_0$ of size $N$ via LHS\;
    \end{itemize}

    \For{$t = 1$ \KwTo $T$}{
        \ForEach{$\mathcal{X} \in P_t$}{
            Evaluate objectives using surrogate models: 
            \[
            \bm{f}(\mathcal{X}) = \left[S(\mathcal{X}), \sigma(\mathcal{X}), \frac{1}{\kappa(\mathcal{X})}\right]^T
            \]
        }

        \textbf{Step 2: Non-dominated Sorting} \\
        Partition $P_t$ into Pareto-fronts $\mathcal{F}_1, \mathcal{F}_2, ...$\;

        \textbf{Step 3: Reference Direction Association} \\
        Assign each $\mathcal{X}$ to the closest $\bm{w}_h$ based on perpendicular distance:
        \[
        d(\bm{f}(\mathcal{X}), \bm{w}_h) = \left\| \bm{f}(\mathcal{X}) - \frac{\bm{f}(\mathcal{X})^T \bm{w}_h}{\|\bm{w}_h\|^2} \bm{w}_h \right\|
        \]

        \textbf{Step 4: Environmental Selection} \\
        Construct $P_{t+1}$ by selecting from $\mathcal{F}_1, \mathcal{F}_2, ...$. If a front exceeds capacity, prioritize solutions maximizing:
        \[
        \max_{\mathcal{X}^*} \left(S(\mathcal{X}^*), \sigma(\mathcal{X}^*), \frac{1}{\kappa(\mathcal{X}^*)}\right)
        \]
        while maintaining diversity across $\bm{w}_h$\;

        \textbf{Step 5: Variation}
        \begin{itemize}
            \item Select parents via tournament selection\;
            \item Apply SBX ($p_{\rm c}=0.9$) for crossover\;
            \item Apply polynomial mutation ($\eta=20$) for diversity\;
            \item Generate offspring population $Q_t$\;
        \end{itemize}
        \textbf{Step 6: Population Update} \\
        Merge parent and offspring populations:
        \[
        P_{t+1} = \text{EnvironmentalSelection}(P_t \cup Q_t, N)
        \]
    }
    \Return $\mathcal{P}^* = \mathcal{F}_1$, $\mathcal{X}^*$\;
\end{algorithm}

\subsection{Details about WaveTENet}

For regression problems, the structure of a dataset $\mathcal{I}$ with \(N\) samples can be summarized as:
\begin{align}
    \mathcal{I}=\left\{\mathcal{X}_i,\left(y_i^S,y_i^{\sigma},y_i^{\kappa}\right) \mid 1 \leq i \leq N\right\}
\end{align}

For the sake of integration, we incorporate the feature generation and wavelet transform modules into the WaveTENet network, as outlined in Algorithm~\ref{alg:waveTENet_feature}.

\begin{algorithm}[H]
    \caption{WaveTENet Feature Construction}
    \label{alg:waveTENet_feature}
    \KwIn{Chemical formula $\mathcal{F}$, e.g., Mg$_{3.5}$Ho$_{0.01}$Sb$_2$}
    \KwOut{Normalized feature vector $\mathcal{X}$}

    \textbf{Step 1: Generate Descriptors} \\
    \Indp Extract SIMD descriptor:
    \[
    \mathcal{S} = \text{SIMD}(\mathcal{F})
    \]
    Extract Magpie descriptor:
    \[
    \mathcal{M} = \text{Magpie}(\mathcal{F})
    \]

    Concatenate SIMD and Magpie descriptors:
    \[
    \mathcal{V} = \mathcal{S} \oplus \mathcal{M}
    \]
    \Indm

    \textbf{Step 2: Apply Haar Wavelet Transform} \\
    \Indp Compute high-frequency coefficients:
    \[
    \mathcal{D} = \text{HaarHigh}(\mathcal{V})
    \]
    Compute low-frequency coefficients:
    \[
    \mathcal{A} = \text{HaarLow}(\mathcal{V})
    \]
    \Indm

    \textbf{Step 3: Concatenate All Features} \\
    \[
    \mathcal{X} = \mathcal{S} \oplus \mathcal{M} \oplus \mathcal{A} \oplus \mathcal{D}
    \]

    \textbf{Step 4: Normalize Feature Vector} \\
    \[
    \mathcal{X} = \text{MinMaxNormalize}(\mathcal{X})
    \]

    \Return $\mathcal{X}$
\end{algorithm}

The objective of the WaveTENet is to minimize the error function to fit the mapping between features and targets, a process typically achieved through backpropagation. Additionally, to prevent overfitting, \(\ell_2\) regularization is applied to enhance model robustness. The error function of WaveTENet can be expressed as:
\begin{align}
    \mathcal{L} = \frac{1}{N} \sum_{i=1}^{N} \left[y_i - f(\mathcal{X}_i;\theta)\right]^2 + \lambda \sum_{j} \|\theta_j\|^2
\end{align}
where \( f(\mathcal{X}_i;\theta) \) represents the model's prediction given parameters \( \theta \) and input features \( \mathcal{X}_i \), \( \lambda \) is the decay coefficient for \( \ell_2 \) regularization, and \( \sum_{j} \|\theta_j\|^2 \) denotes the sum of the squared \( \ell_2 \) norms of all model parameters. Algorithm~\ref{alg:waveTENet_bp} describes the forward and backward propagation of WaveTENet to achieve target prediction and parameter optimization, respectively.

\begin{algorithm}[H]
    \caption{WaveTENet Backpropagation}
    \label{alg:waveTENet_bp}
    \KwIn{Training data $(\mathcal{X}, y)$, learning rate $\alpha$, weight decay $\lambda$, number of epochs $T$}
    \KwOut{Optimized model parameters $\mathcal{\theta}$}

    \For{$t = 1$ \KwTo $T$}{
        \textbf{Step 1: Forward Pass} \\
        \ForEach{$(\mathcal{X}_i, y_i) \in (\mathcal{X}, y)$}{
            Compute input transformation:
            \[
            \mathcal{H}^{(0)} = \text{ReLU}(\text{BN}(\mathcal{W}_0 \mathcal{X}_i + \bm{b}_0))
            \]

            \For{$l = 1$ \KwTo $L$}{ 
                Compute dense block output:
                \[
                \mathcal{H}^{(l)} = \text{ReLU}(\text{Dropout}(\text{BN}(\mathcal{W}_l \mathcal{H}^{(l-1)} + \bm{b}_l)))
                \]
                Residual connection:
                \[
                \mathcal{H}^{(l)} = \mathcal{H}^{(l)} + \mathcal{H}^{(l-1)}
                \]
            }

            Compute final output:
            \[
            \hat{y}_i = \mathcal{W}_{\text{out}} \mathcal{H}^{(L)} + \bm{b}_{\text{out}}
            \]
        }

        \textbf{Step 2: Compute Loss} \\
        \[
        \mathcal{L} = \frac{1}{N} \sum_{i=1}^{N} \left[y_i - \hat{y}_i \right]^2 + \lambda \sum_{j} \|\mathcal{\theta}_j\|^2
        \]

        \textbf{Step 3: Backpropagation} \\
        \ForEach{layer $l$ from $L$ to $0$}{
            Compute gradients:
            \[
            \frac{\partial \mathcal{L}}{\partial \mathcal{W}_l} = \frac{1}{N} \sum_{i=1}^{N} 2(y_i - \hat{y}_i)(-\mathcal{H}^{(l)}) + \lambda \mathcal{W}_l
            \]
            \[
            \frac{\partial \mathcal{L}}{\partial \bm{b}_l} = \frac{1}{N} \sum_{i=1}^{N} 2(y_i - \hat{y}_i)(-1)
            \]
        }

        \textbf{Step 4: Update Parameters} \\
        \ForEach{parameter $\mathcal{\theta}_j \in \{\mathcal{W}_l, \bm{b}_l\}$}{
            \[
            \mathcal{\theta}_j = \mathcal{\theta}_j - \alpha \frac{\partial \mathcal{L}}{\partial \mathcal{\theta}_j}
            \]
        }
    }

    \Return Optimized parameters $\mathcal{\theta}$
\end{algorithm}

\section{Data Availability}
The ESTM (for models training) and UCSB (for transfer learning) datasets can be accessed via our GitHub repository at \url{https://github.com/FlorianTseng/WaveTENet}.

\section{Code Availability}
\label{sec:code_avail}
The codes supporting our research are available at \url{https://github.com/FlorianTseng/WaveTENet}.

\section{Competing Interests}
The authors declare that they have no known competing financial interests or personal relationships that could have appeared to influence the work reported in this paper.

\begin{acknowledgments}
We would like to acknowledge the National Key Research and Development Program of China (Grant No. 2023YFB4603800), the financial support from the project of the National Natural Science Foundation of China (Grants No.12404062, No.12474093, No.12302220, No.12374091), Translational Medicine and Interdisciplinary Research Joint Fund of Zhongnan Hospital of Wuhan University (Grant NO.ZNJC202235), and the Fundamental Research Funds for the Central Universities (Grant No.2042023kf0109). The numerical calculations in this work have been done on the supercomputing system in the Supercomputing Center of Wuhan University.
\end{acknowledgments}

\bibliography{ref}
\bibliographystyle{naturemag}

\end{document}